\documentclass[12pt]{article} 
\usepackage{amssymb,amsmath,color,graphicx,float}

\numberwithin{equation}{section}
\usepackage[toc,page]{appendix}

\newcommand{\nc}{\newcommand}

\nc{\la}{\lambda} \nc{\alf}{\alpha} \nc{\La}{\Lambda} \nc{\ze}{\zeta}
\nc{\tht}{\theta} \nc{\T}{\Theta} \nc{\be}{\beta}  \nc{\eps}{\epsilon} 
\nc{\ga}{\gamma}  \nc{\De}{\Delta}  \nc{\G}{\Gamma}  \nc{\vphi}{\varphi}
\nc{\de}{\delta} \nc{\si}{\sigma}  \nc{\ka}{\kappa}   \nc{\Si}{\Sigma} 
\nc{\om}{\omega}  \nc{\qq}{\qquad}                \nc{\Om}{\Omega} \nc{\vrho}{\varrho}
\nc{\nf}{\infty}   \nc{\dl}{\mathop{\smash{\cal L}}}  \nc{\black}{\rule{3mm}{3mm}}
\nc{\ra}{\rightarrow}    \nc{\ol}{\overline}        \nc{\und}{\underline} 
\nc{\beq}{\begin{equation}}  \nc{\eeq}{\end{equation}}  \nc{\pt}{\partial}  
   \nc{\dst}{\displaystyle}  \nc{\na}{\nabla} 
\nc{\nnb}{\nonumber}    \nc{\bs}{\backslash}        \nc{\mb}{\mathbb}   
\nc{\sn}{{\rm sn}\,} \nc{\cn}{{\rm cn}\,}     \nc{\dn}{{\rm dn}\,} \nc{\nin}{\noindent}
\nc{\ti}{\tilde}   \nc{\wti}{\widetilde}   \nc{\h}{\hat}  \nc{\wh}{\widehat}
\nc{\tpsi}{\wti{\psi}}   \nc{\tphi}{\wti{\phi}}  \nc{\tH}{\wti{H}} \nc{\Ai}{{\rm Ai}}
\nc{\Pf}{P_{\phi}}  \nc{\Pt}{P_{\tht}}

\newcounter{muni}
\newenvironment{remunerate}{\begin{list}{{\rm \arabic{muni}.}}
{\usecounter{muni}
\setlength{\leftmargin}{0pt}\setlength{\itemindent}{38pt}}}{\end{list}}

\nc{\brm}{\begin{remunerate}}   \nc{\erm}{\end{remunerate}}

 \newtheorem{nlem}{Lemma}
\newtheorem{nth}{Proposition}  \newtheorem{nTh}{Theorem}
\newtheorem{ndef}{Definition}     

\nc{\stg}{\mathop{\smash{*}}}
\nc{\st}{\mathop{\smash{\delta}}}
\nc{\barr}{\begin{array}}   \nc{\earr}{\end{array}}   \nc{\dg}{\dagger}
\nc{\mtvb}{\mathversion{bold}}   \nc{\mtvn}{\mathversion{normal}} 

\begin{document}

\date{today}

\begin{titlepage}

\vskip 0.5truecm\centerline{\Large\bf Superintegrable models versus}

\vskip 0.5truecm
\centerline{\Large\bf Zoll metrics of revolution}

\vskip 0.5truecm

\vskip 1.0truecm
\centerline{ \large\bf Galliano VALENT}  

\vspace{1cm}
\centerline{LPMP: Laboratoire de Physique Math\'ematique de Provence}

\vspace{3mm} 
\centerline{13100 Aix en Provence, France.} 

\vspace{1cm}
\centerline{\em Dedicated to the memory of Christian Duval (1947-2018)}

\vspace{3mm}
\centerline{``Entre toutes les passions de l'esprit humain,} 
\centerline{l'une des plus violentes, c'est le d\'esir de savoir"}

\vspace{5mm} \centerline{25/02/2020}

\vskip 1.5truecm

\begin{abstract} Koenigs constructed a family of two dimensional superintegrable (SI) models with one linear and two quadratic integrals in the momenta, shortly $(1,2)$. More recently Matveev and Shevchishin have shown that this construction does generalize to models with one linear and two cubic integrals i.e. $(1,3)$, up to the solution of a non-linear ordinary differential equation. Our explicit solution of this equation allowed for the construction of these SI systems and led to the proof that the systems globally defined on ${\mb S}^2$ are Zoll. We will generalize these results to the case 
$(1,n)$ for any  $n\geq 2$. Our approach is again constructive and shows the existence, when $n$ is odd, of  metrics globally defined on ${\mb S}^2$ which are indeed Zoll (under appropriate restrictions on the parameters), while if $n$ is even the metrics we found are never globally defined on ${\mb S}^2$, as it is already the case for the $(1,2)$ models constructed by Koenigs.
\end{abstract}

\vspace{2cm}

\noindent MSC (2010): {\tt 32C05}, {\tt 53C22}, 
{\tt 37E99}, {\tt 37J35}, {\tt  37K25}, {\tt 81V99}.

\end{titlepage}

\newpage
\section{Introduction}
As explained in the abstract, the starting point of our work is a set of SI models due to Koenigs \cite{Ko}, as popularized and generalized in \cite{kk1},\cite{kk2}. These models, defined on surfaces of revolution, exhibit  an hamiltonian with one linear and two quadratic integrals in the momenta. Let us give an example, using the coordinates of \cite{Va1}, with hamiltonian
\beq\label{Koex}
H=\frac{\cosh^2 x}{2(\rho+\sinh x)}(P_x^2+\Pf^2).
\eeq
The symmetry of revolution shows that $(H,\Pf)$ is already an integrable system. To reach SI we need a set of extra integrals
\beq
S_1=\cos\phi\,(H-\sinh x\,\Pf^2)+\sin\phi\,(\cosh x\,P_x\,\Pf), \qq S_2=\{\Pf,S_1\}.
\eeq
The extra integrals are not algebraically independent since we have 
\beq\label{square}
S_1^2+S_2^2=H^2+2\rho\,H\,\Pf^2-\Pf^4.\eeq
However, the main problem, as pointed out in \cite{Va1},  is that the metric (\ref{Koex}) is never globally defined on the manifold ${\mb S}^2$. This unpleasant feature led Matveev and Shevchishin to take cubic extra integrals rather than quadratic ones. Still considering a  surface of revolution
\beq
H=\Pi^2+h_x^2\,\Pf^2 \qq\qq \Pi=h_x\,P_x,\qq\qq 
h_x=\frac{dh}{dx},
\eeq
they started from
\beq
S_1=\cos\phi\,\Pi\,(H+\la_1(x)\,\Pf^2)+\sin\phi\,\Pf(\la_0(x)\,H+\la_2(x)\,\Pf^2),\qq S_2=\{\Pf,S_1\}.
\eeq
Here too one gets
\beq
S_1^2+S_2^2=H^3+\si_1\,H^2\,\Pf^2+\si_2\,H\,\Pf^4+\si_3\,\Pf^6,\eeq
with appropriate constants $\si_i$.

However, Matveev and Shevchishin were led to a non-linear first order ODE which they could not solve. It was solved in \cite {vds} through appropriate coordinates changes and stemmed with the discovery of a metric (exhibiting two parameters) globally defined on ${\mb S}^2$. In a subsequent work \cite {Va2} we proved that this family of metrics is indeed Zoll.

We were led to consider the general case where the extra integrals $S_1$ and $S_2$ are of any integer degree $n$ in the momenta, starting from Koenigs for $n=2$. In fact, to go through, the analysis needs to consider separately the odd and the even degrees.

The plan of this article is the following. In Section 2 we state our results in two Theorems, dealing successively with the case of extra integrals of degree $2n+1$ for $n\geq  1$, and extra integrals of degree $2n$ for $n\geq 1$. Then Section 3 gives the proof of Theorem 1. In Section 4 the geodesics are constructed on the one hand using action-angle coordinates and on the other hand using the extra integrals. In Section 5 is given the proof of Theorem 2.  Some concluding remarks are presented in the Section 6.


\section{The results}
When looking for a surface of revolution on 
${\mb S}^2$,  as shown in \cite{Be}[Proposition 4.10], one may start with the metric
\beq
g=A^2(\tht)\,d\tht^2+\sin^2\tht\,d\phi^2\qq\qq \tht\in\,(0,\pi)\qq \phi\in\,{\mb S}^1,
\eeq
leading to the hamiltonian 
\beq\label{ham1}
H=\Pi^2+\frac{\Pf^2}{\sin^2\tht}\qq\qq \Pi=\frac{\Pt}{A(\tht)}.
\eeq
The Killing vector $\pt_{\phi}$ implies, at the hamiltonian level, the conservation of $\Pf$. In such a way the pair $(H,\,\Pf)$ already defines an integrable system. To switch to a SI one, let us add two {\em extra} integrals 
\beq\label{extraI}
S_1=\cos\phi\,{\cal S}+\sin\phi\,{\cal T},\qq S_2=\{\Pf,S_1\}=\cos\phi\,{\cal T}-\sin\phi\,{\cal S},
\eeq
where ${\cal S}$ and ${\cal T}$ are polynomials in $H$ and in $\Pf^2$, of fixed degree in the momenta, denoted by $\sharp({\cal S})=\sharp({\cal T})$.

Our first result is:
\begin{nTh} In the case where $\sharp(S_1)=\sharp(S_2)=2n+1$ with $n\geq 1$, the system defined by
\beq
H,\qq \Pf, \qq S_1, \qq S_2,\eeq
where the extra integrals (\ref{extraI}) are built with
\beq
{\cal S}=\Pi\sum_{k=0}^n\,\la_{2k-1}(\tht)\,H^{n-k}\,\Pf^{2k}, \qq\qq {\cal T}=\sum_{k=0}^n\,\la_{2k}(\tht)\,H^{n-k}\,\Pf^{2k+1},
\eeq  
is superintegrable if one takes
\beq
A(\tht)=1+\cos\tht\,\sum_{k=1}^{2n}\,\frac{e_k}{\sqrt{1-m_k\,\sin^2\tht}}, \qq\quad \forall k: \quad e_k^2=1, 
\eeq
where all of the $2n$ real parameters $m_k$ are restricted to $m_k<1$. 

If, in addition, we have 
\beq   \sum_{k=1}^{2n}\,e_k=0, \eeq
and \footnote{For $n=1$ this restriction is not required.}
\beq
{\cal A}_0^{(n)}\equiv \sum_{k=1}^{n}\,\left|1-\sqrt{\frac{\mu_{2k-1}}{\mu_{2k}}}\right|<1, \qq\quad  (\mu_k=1-m_k),
\eeq
then the SI system is globally defined on ${\mb S}^2$ and the metric is Zoll. 

The set of observables
\[
H,\qq \Pf, \qq S_+=S_1+iS_2, \qq S_-=S_1-iS_2\]
generates a Poisson algebra, with the relations \footnote{The relation between the $\si_l$ and the parameters $m_k$ is given in Proposition \ref{compsigma}.}
\beq
S_+\,S_-=\sum_{l=0}^{2n+1}\,\si_l\,H^{2n+1-l}\,\Pf^{2l}
\eeq
and
\beq
\{S_+,S_-\}=2i\sum_{l=0}^{2n}\,(l+1)\si_{l+1}\,H^{2n-l}\,\Pf^{2l+1}.\eeq
\end{nTh}

Our second result is:

\begin{nTh} In the case where $\sharp(S_1)=\sharp(S_2)=2n$ with $n\geq 1$, the system defined by
\beq
H,\qq \Pf, \qq S_1, \qq S_2,
\eeq
where the extra integrals (\ref{extraI}) are built with
\beq\label{AB2}
{\cal S}=\sum_{k=0}^n\,\la_{2k-1}(\tht)\,H^{n-k}\,\Pf^{2k}, \qq\qq {\cal T}=\Pi\sum_{k=0}^{n-1}\,\la_{2k}(\tht)\,H^{n-k-1}\,\Pf^{2k+1},
\eeq  
is SI if one takes 
\beq
A(\tht)=1+\cos\tht\,\sum_{k=1}^{2n-1}\,\frac{e_k}{\sqrt{1-m_k\,\sin^2\tht}}, \qq\qq  e_k^2=1, 
\eeq
and all of the $2n-1$ real parameters $m_k$ are restricted to $m_k<1$.  

This system is {\bf never} globally defined on 
${\mb S}^2$. 
\end{nTh}
Let us begin with the proof of Theorem 1.


\section{Integrals of odd degree in the momenta}
Here $\sharp(S_1)=\sharp(S_2)=2n+1$ for $n\geq 1$. 
Let us recall that the hamiltonian is given by
\beq\label{hamimpair}
H=\Pi^2+\sin^2\tht\,\Pf^2,\qq\qq\Pi=\frac{\Pt}{A(\tht)},
\eeq
and the extra integrals by
\beq
S_1=\cos\phi\,{\cal S}+\sin\phi\,{\cal T},\qq S_2=\{\Pf,S_1\}=\cos\phi\,{\cal T}-\sin\phi\,{\cal S},
\eeq
where
\beq\label{AB1}
{\cal S}=\Pi\sum_{k=0}^n\,\la_{2k-1}(\tht)\,H^{n-k}\,\Pf^{2k}, \qq\qq {\cal T}=\sum_{k=0}^n\,\la_{2k}(\tht)\,H^{n-k}\,\Pf^{2k+1}.
\eeq  
These integrals are therefore defined by an array of functions of 
$\tht$ of the form
\[\left(\barr{ccccc}
\la_{-1}=1 &\quad \la_1 & \quad \la_3 &\quad \ldots & \la_{2n-1}\\[4mm]
\la_0 & \quad \la_2 & \quad \la_4 & \quad\ldots & \quad \la_{2n}\earr\right),\]
provided that they are determined by

\begin{nth}\label{sd11} $S_1$ and $S_2$ will be integrals iff the $\la$'s s solve the differential system \footnote{A prime is a $\tht$ derivative while $s=\sin\tht$ and $c=\cos\tht$.}:
\beq\label{sd1}0\leq k \leq n: \qq 
\left\{\barr{cclc}
s^2\,\la'_{2k} & = & A\,\la_{2k-1},
& \qq (a)\\[4mm]
s^2\,\la'_{2k+1} & = & \dst \la'_{2k-1}-\frac cs\,\la_{2k-1}-A\,\la_{2k}, & \qq (b)
\earr\right.\eeq
with the conventional value  $\la_{2n+1}=0$.
\end{nth}

\vspace{5mm}\nin{\bf Proof:} Both constraints $\{H,S_1\}=0$ and $\{H,S_2\}=0$ are seen to be equivalent to
\beq
\{H,{\cal S}\}=-2\frac{\Pf}{s^2}\,{\cal T} \qq\qq 
\{H,{\cal T}\}=2\frac{\Pf}{s^2}\,{\cal S}.
\eeq
Using the explicit form of ${\cal S}$ and ${\cal T}$ elementary computations give (\ref{sd1}).$\hfill\Box$

\vspace{5mm}\nin{\bf Remark:} 
One can get rid of the derivatives in the right hand side of relation (\ref{sd1}) by a simple recurrence which gives for $0\leq k\leq n-1$:
\beq
s^{2(k+1)}\,\la'_{2k+1}=-\frac cs(1+s^2\la_1+s^4\la_3+\ldots+s^{2k}\la_{2k-1})-(\la_0+s^2\la_2+\ldots+s^{2k}\la_{2k})A,
\eeq
while for $k=n$ one gets the purely algebraic relation
\beq\label{alg1}
\frac cs (1+s^2\la_1+s^4\la_3+\ldots+s^{2n}\la_{2n-1})=-
(\la_0+s^2\la_2+\ldots+s^{2n}\la_{2n})A.
\eeq

A simplifying approach to the differential system (\ref{sd1}) makes use of generating functions, which encode all of the $\la$'s in a couple of objects.

\begin{ndef} Let us define the generating functions
\beq\label{defgf1}
{\cal L}(\tht,\xi)=\sum_{k=0}^n\,\la_{2k}(\tht)\,\xi^k, \qq\qq 
{\cal M}(\tht,\xi)=\sum_{k=0}^n\la_{2k-1}(\tht)\,\xi^k,
\qq\quad \xi\in{\mb C}.
\eeq\end{ndef}
These objects are mere polynomials in the variable $\xi$. 
Their usefulness follows from
\begin{nth}\label{odegf} The differential system (\ref{sd1}) is equivalent, in terms of the generating functions, to
\beq\label{eqgf1}
s^2\,\pt_{\tht}{\cal L}=A\,{\cal M},\qq\qq  
s^2(1+\tau)\pt_{\tht}{\cal M}+\xi\,\frac cs\,{\cal M}=-\xi\,A\,{\cal L}, \qq\qq \tau=-\frac{\xi}{s^2}.\eeq
\end{nth}

\vspace{5mm}
\nin{\bf Proof:} Upon use of relations (a) in (\ref{sd1}) we have
\beq
s^2\,\pt_{\tht}{\cal L}(\tht,\xi)=\sum_{k=0}^n\xi^k\,s^2\,\la'_{2k}(\tht)=A\sum_{k=0}^n\xi^k\,\la_{2k-1}(\tht)=A\,{\cal M}(\tht,\xi).
\eeq
Conversely, expanding this relation in powers of $\xi$ gives back the relations (a) in (\ref{sd1}).

Similarly, using relations (b) in (\ref{sd1}) we get
\beq
s^2\pt_{\tht}{\cal M}=\sum_{k=1}^n\xi^k\,s^2\,\la'_{2k-1}=\sum_{k=1}^n\xi^k\Big(\la'_{2k-3}-\frac cs\,\la_{2k-3}-A\,\la_{2k-2}\Big),
\eeq
which becomes
\beq
s^2\pt_{\tht}{\cal M}=\pt_{\tht}\Big(\xi\,{\cal M}-\xi^{n+1}\la_{2n-1}\Big)-\frac cs\Big(\xi{\cal M}-\xi^{n+1}\la_{2n-1}\Big)-A\Big(\xi{\cal L}-\xi^{n+1}\la_{2n}\Big).
\eeq
We end up with
\beq
s^2(1+\tau)\pt_{\tht}{\cal M}+\xi\frac cs\,{\cal M}+\xi\,A\,{\cal L}=-\xi^{n+1}(\pt_{\tht}\la_{2n-1}-\la_{2n-1}-A\,\la_{2n}),\eeq
and the right hand member does vanish thanks to relation (b)  for $k=n$ in (\ref{sd1}). Conversely, expanding this relation in powers of $\xi$ one recovers relations (b) in (\ref{sd1}). $\hfill\Box$

Let us describe the structure of the array of the 
$\la_k$.

\subsection{The solution for integrals of odd degree}
Let us define the functions
\beq\label{hk}
\forall k\in\{1,2,\ldots,2n\}: \qq h_k(\tht)=e_k\,\sqrt{1-m_k\,s^2},\qq e_k^2=1, \qq m_k<1,\eeq
and
\beq\label{gfH}
{\cal H}(\tht,\xi)\equiv \prod_{k=1}^{2n}(1+\xi\,h_k(\tht))=\sum_{k=0}^{2n}\,(H)_k(\tht)\,\xi^k.
\eeq
The $(H)_k(\tht)$ are nothing but the symmetric functions constructed in terms of the $h_k(\tht)$. Their explicit form is
\beq
(H)_0(\tht)=1 \qq\qq (H)_1(\tht)=\sum_{l=1}^{2n}\,h_l(\tht), 
\eeq
and more generally
\beq
(H)_k(\tht)=\sum_{1\leq l_1 \leq l_2\leq \ldots\leq l_k \leq 2n}\,h_{l_1}(\tht)\,h_{l_2}(\tht)\cdots h_{l_k}(\tht), \qq\qq 2\leq k\leq 2n.
\eeq
In terms of these objects, we can write down the solution for the $\la$'s. 

\begin{ndef}\label{sollambda} Let us define, for $k\in\{1,2,\ldots,n\},$ the functions
\beq\label{laimpair}
\la_{2k-1}=\frac{(-1)^k}{s^{2k}}\left\{\sum_{l=0}^k\,{n-l \choose n-k}\,(H)_{2l}+c\,\sum_{l=0}^{k-1}\,{n-l-1  \choose n-k}\,(H)_{2l+1}\right\},
\eeq
and for $k\in\{1,2,\ldots,n-1\}$:
\beq\label{lapair}
\quad \la_{2k}=\frac{(-1)^{k+1}}{s^{2k+1}}\left\{\sum_{l=0}^k\,{n-l \choose n-k}\,(H)_{2l+1}+c\sum_{l=0}^k\,{n-l \choose n-k}\,(H)_{2l}\right\},\eeq
as well as
\beq\label{laspe1}
\la_{2n}=\frac{(-1)^{n+1}}{s^{2n+1}}\left\{\sum_{l=0}^{n-1}\,(H)_{2l+1}+c\sum_{l=0}^n\,(H)_{2l}\right\}.
\eeq

\end{ndef}
A direct proof that these formulae do solve the differential system (\ref{sd1}) is rather cumbersome. We will first compute their generating functions and then use Proposition \ref{odegf}.

\begin{nth} The generating functions ${\cal L}$ and 
${\cal M}$ are given by
\beq\label{gfL}
-s\,{\cal L}(\tht,\xi)=\sum_{l=0}^{n-1}\,\psi_{l,n}\,(H)_{2l+1}+c\,\sum_{l=0}^n\,\psi_{l,n}\,(H)_{2l}
\eeq
and by
\beq\label{gfM}
{\cal M}(\tht,\xi)=\sum_{l=0}^n\,\psi_{l,n}\,(H)_{2l}+c\,\sum_{l=0}^{n-1}\,\psi_{l+1,n}\,(H)_{2l+1}
\eeq
where
\beq
\tau=-\frac{\xi}{s^2}, \qq\qq \psi_{l,n}=\tau^l(1+\tau)^{n-l}, \qq  0\leq l \leq n.
\eeq
\end{nth}

\vspace{5mm}
\nin{\bf Proof:} Starting from
\beq
-s\,{\cal L}(\tht,\xi)=\sum_{k=0}^{n-1}\,\xi^k\,(-s\,\la_{2k})+\xi^n\,(-s\,\la_{2n}),
\eeq
using (\ref{lapair}),(\ref{laspe1}) and interchanging the order of the summations we get 
\beq 
\sum_{l=0}^{n-1}\,(H)_{2l+1}\sum_{k=l}^{n-1}{n-l \choose n-k}\tau^{k}+c\,\sum_{l=0}^{n-1}\,(H)_{2l}\sum_{k=l}^{n-1}{n-l \choose n-k}\tau^{k},
\eeq
to which we must add
\beq
\xi^n\,(-s\,\la_{2n})=\sum_{l=0}^{n-1}\,(H)_{2l+1}\,\tau^n+c\,\sum_{l=0}^n\,(H)_{2l}\,\tau^n,\eeq
leading to
\beq
-s\,{\cal L}(\tht,\xi)=\sum_{l=0}^{n-1}\tau^l\Big((H)_{2l+1}+c\,(H)_{2l}\Big)\sum_{k=l}^n{n-l \choose k-l}\tau^{k-l}+c\,\tau^n\,(H)_{2n}.
\eeq
Using the binomial theorem we get
\beq
-s\,{\cal L}(\tht,\xi)=\sum_{l=0}^{n-1}\tau^l(1+\tau)^{n-l}\Big((H)_{2l+1}+c\,(H)_{2l}\Big)+c\,\tau^n\,(H)_{2n},
\eeq
leading to (\ref{gfL}).

For ${\cal M}$, using (\ref{laimpair}), we have
\beq
{\cal M}(\tht,\xi)=\sum_{k=0}^n\tau^k\sum_{l=0}^k\,{n-l \choose n-k}(H)_{2l}+c\sum_{k=1}^n\tau^k\sum_{l=0}^{k-1}{n-l-1 \choose n-k}(H)_{2l+1}.
\eeq
Interchanging the summation order and using the binomial theorem gives for the first sum
\beq
\sum_{l=0}^n\tau^l\,(H)_{2l}\sum_{k=l}^n {n-l \choose k-l}\tau^{k-l}=\sum_{l=1}^n\tau^l(1+\tau)^{n-l}\,(H)_{2l}
\eeq
and, by the same token, for the second one
\beq\barr{l}\dst 
c\sum_{k=1}^n\tau^k\sum_{l=0}^{k-1}{n-l-1 \choose n-k}(H)_{2l+1}=c\sum_{l=0}^{n-1}\tau^l\,(H)_{2l+1}\sum_{k=l+1}^n{n-l-1 \choose k-l-1}\tau^{k-l}=\\[4mm]\dst 
\hspace{4cm} =c\sum_{l=0}^{n-1}\tau^{l+1}(1+\tau)^{n-l-1}\,(H)_{2l+1}.\earr
\eeq
This concludes the proof. $\hfill\Box$

\vspace{5mm}\nin{\bf Remark:} Using this Proposition one  can check relation (\ref{alg1}) which becomes, in terms of the generating functions:
\beq
\left(\frac cs\,{\cal M}+A\,{\cal L}\right)
\Big|_{\tau=-1}=0.
\eeq

Up to now we have defined our $\la_i$ and computed their generating functions. We reach the core of this first part: we will prove the PDE's for the generating functions which determine the explicit form of the function $A(\tht)$ and of the hamiltonian.

\begin{nth}\label{GF1} The generating functions given by (\ref{gfL}) and by (\ref{gfM}) are solutions of the following  equations 
\beq\label{eq2et31} 
s^2\,\pt_{\tht}{\cal L}=A\,{\cal M}\qq\qq  s^2\,(1+\tau)\,\pt_{\tht}{\cal M}+\xi\frac cs\,{\cal M}=-\xi\,A\,{\cal L},\eeq
where
\beq
A(\tht)=1+c\sum_{k=1}^{2n}\frac{e_k}{\sqrt{1-m_k\,s^2}}.
\eeq
It follows that the $\la$'s given by  Definition \ref{sollambda} are indeed a solution of the differential system (\ref{sd1}), and this implies that $S_1$ and $S_2$ are integrals for the hamiltonian 
\beq
H=\Pi^2+\frac{\Pf^2}{s^2},\qq\quad \Pi=\frac{\Pt}{A(\tht)}.\eeq
\end{nth} 
 
\vspace{5mm}\nin{\bf Proof:} Let us first define the following splitting of the generating functions:
\beq
{\cal L}={\cal L}_1+{\cal L}_2\eeq
where
\beq
{\cal L}_1=-\frac 1s\sum_{l=0}^{n-1}\psi_{l,n}(H)_{2l+1} \qq \qq{\cal L}_2=-\frac cs\sum_{l=0}^n\psi_{l,n}(H)_{2l},\eeq
and similarly
\beq
{\cal M}={\cal M}_1+{\cal M}_2\eeq
where
\beq
{\cal M}_1=\sum_{l=0}^{n}\psi_{l,n}(H)_{2l} \qq\qq 
{\cal L}_2=c\sum_{l=0}^{n-1}\psi_{l+1,n}(H)_{2l+1}.
\eeq

Let us compute first $s^2\pt_{\tht}{\cal L}_1$. It is made out of two pieces. The first one, which follows from:
\beq
-s\pt_{\tht}\psi_{l,n}=2c\,\tau\pt_{\tau}\,\psi_{l,n}=c\Big(2l\,\psi_{l,n}+2(n-l)\,\psi_{l+1,n}\Big),
\eeq
is given by
\beq
c\sum_{l=0}^{n-1}\,(2l+1)\psi_{l,n}(H)_{2l+1}+c\sum_{l=0}^{n-2}2(n-l)\psi_{l+1,n}(H)_{2l+1}+2c\,\psi_{n,n}(H)_{2n-1}.\eeq
The second piece follows from relation (\ref{derH}) in Appendix A \footnote{Recall that here $\nu=2n$.}
\beq\label{Himpair}
-s\pt_{\tht}\,(H)_{2l+1}=-c\Big((2l+1)\,(H)_{2l+1}+(2n-2l+1)\,(H)_{2l-1}\Big)+(A-1)\,(H)_{2l},
\eeq
and is given by
\beq
-c\sum_{l=0}^{n-1}\,(2l+1)\psi_{l,n}(H)_{2l+1}-\sum_{l=0}^{n-2}(2n-2l+1)\psi_{l+1,n}(H)_{2l+1}+(A-1)\sum_{l=0}^{n-1}\psi_{l,n}(H)_{2l}.\eeq
Adding up we are left with
\beq
\sum_{l=0}^{n-1}\psi_{l+1,n}(H)_{2l+1}+(A-1)\sum_{l=0}^{n}\psi_{l,n}(H)_{2l}+\psi_{n,n}\Big(c(H)_{2n-1}-(A-1)(H)_{2n}\Big).\eeq
The last piece vanishes upon use of (\ref{idH}). The final result is
\beq
s^2\pt_{\tht}{\cal L}_1={\cal M}_2+(A-1){\cal M}_1.
\eeq
Similarly one can show
\beq
s^2\pt_{\tht}{\cal L}_2={\cal M}_1+(A-1){\cal M}_2.
\eeq
Adding up we get the first relation in (\ref{eq2et31}).

Let us now compute $s^2(1+\tau)\pt_{\tht}{\cal M}_1$. It is made out of two pieces. The first one, which follows from
\beq
s^2(1+\tau)\pt_{\tht}\,\psi_{l,n}=\xi\frac cs\Big(2l\,\psi_{l-1,n}+2(n-l)\psi_{l,n}\Big), \qq s^2(1+\tau)\psi_{l,n}=-\xi\psi_{l-1,n},
\eeq
is given by
\beq\xi\frac cs\sum_{l=1}^n 2l\,\psi_{l-1,n}(H)_{2l}+\xi\frac cs\sum_{l=0}^{n-1}2(n-l)\psi_{l,n}(H)_{2l}.\eeq
The second piece, which follows from (\ref{derH}):
\beq\label{Hpair}
\pt_{\tht}\,(H)_{2l}=\frac cs\Big(2l\,(H)_{2l}+2(n-l+1)(H)_{2(l-1)}\Big)-\frac{A-1}{s}\,(H)_{2l-1},\qq l\geq 1,
\eeq 
is given by
\beq
-\xi\frac cs\sum_{l=1}^n2l\,\psi_{l-1,n}(H)_{2l}-\xi\frac cs\sum_{l=0}^{n-1}2(n-l)\psi_{l,n}(H)_{2l}+\frac{\xi}{s}(A-1)\sum_{l=1}^n\psi_{l-1,n}(H)_{2l-1}.\eeq
Adding these two pieces one gets
\beq
s^2(1+\tau)\pt_{\tht}{\cal M}_1=-\xi(A-1){\cal L}_1.\eeq
Adding to both members $\dst \xi\frac cs{\cal M}_1=-\xi\,{\cal L}_2$ we conclude to
\beq
s^2(1+\tau)\pt_{\tht}{\cal M}_1+\xi\frac cs{\cal M}_1=\xi({\cal L}_1-{\cal L}_2)-\xi\,A\,{\cal L}_1.\eeq

Similarly one can prove 
\beq
s^2(1+\tau)\pt_{\tht}{\cal M}_2+\xi\frac cs{\cal M}_2=-\xi({\cal L}_1-{\cal L}_2)-\xi\,A\,{\cal L}_2.\eeq
Adding them up we get the second relation in (\ref{eq2et31}). Using Proposition \ref{odegf} we can conclude that $S_1$ and $S_2$ are integrals of $H$.$\hfill\Box$

Having constructed a SI system with a linear integral and two extra integrals of degree $2n+1$ in the momenta, let us show that this solution, under appropriate restrictions on the parameters $m_k$, is globally defined on $M={\mb S}^2$.

\subsection{Global structure}
We have seen that the metric and the hamiltonian
\beq\label{metGV}
g=A^2(\tht)\,d\tht^2+s^2\,d\phi^2, \qq H=\Pi^2+\frac{\Pf^2}{s^2},\qq\qq \Pi=\frac{\Pt}{A(\tht)},
\eeq 
where
\beq\label{metGV1}
A(\tht)=1+{\cal A}(\tht), \quad {\cal A}(\tht)=c\sum_{k=1}^{2n}\frac{e_k}{\sqrt{1-m_k\,s^2}},\qq \forall k:\quad \Big( e_k^2=1 \quad \&\quad m_k<1\Big),
\eeq
exhibits 3 integrals: $(\Pf,\ S_1,\ S_2)$. 

Let us first prove:
\begin{nlem}\label{boundA} If $\dst\sum_{k=1}^{2n}\,e_k=0$ one has the uniform bound 
\beq
\forall\tht \in (0,\pi): \qq |{\cal A}(\tht)|\leq {\cal A}_0^{(n)}\equiv \sum_{k=1}^n\,\left|1-\sqrt{\frac{\mu_{2k-1}}{\mu_{2k}}}\right|, \qq\quad \mu_k=1-m_k>0.\eeq 
\end{nlem}

\vspace{5mm}\nin{\bf Proof:} Since we have $\dst\sum_{k=1}^{2n}\,e_k=0$, we can write
\beq
{\cal A}(\tht)=\sum_{k=1}^{n}\wti{e}_k\left(
\frac c{\sqrt{1-m_{2k-1}\,s^2}}-\frac c{\sqrt{1-m_{2k}\,s^2}}\right), \qq\quad \forall k: \quad \wti{e}_k^2=1. 
\eeq
Since ${\cal A}(\tht)$ is odd, it is sufficient to consider $\tht\in\,[0,\pi/2)$. The substitution $t=\tan\tht$ gives
\beq
{\cal A}(\tht(t))=\sum_{k=1}^n\wti{e}_k\,f_k(t)
\qq f_k(t)=\left(\frac 1{\sqrt{1+\mu_{2k-1}\,t^2}}
-\frac 1{\sqrt{1+\mu_{2k}\,t^2}}\right), \qq t\geq 0.\eeq
Writing
\[
f_k(t)=\frac{(\sqrt{\mu_{2k}}-\sqrt{\mu_{2k-1}})}{\sqrt{\mu_{2k}}}
\,\frac{\sqrt{\mu_{2k}}\,t}{\sqrt{1+\mu_{2k}\,t^2}}\,\frac 1{\sqrt{1+\mu_{2k-1}\,t^2}}
\frac{\sqrt{\mu_{2k}}\,t+\sqrt{\mu_{2k-1}}\,t}{\sqrt{1+\mu_{2k}\,t^2}+\sqrt{1+\mu_{2k-1}\,t^2}},\]
and observing that each term in the product is uniformly bounded, for $t\geq 0$ by $1$, we get:
\[ \forall t \geq 0: \qq |f_k(t)|\leq\left|1-\sqrt{\frac{\mu_{2k-1}}{\mu_{2k}}}\right|,  \]
implying the lemma. $\hfill\Box$

\begin{nth} The SI system of observables
\[ H, \quad \Pf, \quad S_1, \quad S_2\]
constructed in Section 3.1 is globally defined on ${\mb S}^2$ and Zoll if 
\beq\label{cond1}
\forall k: \quad m_k<1, \qq \& \qq\sum_{k=1}^{2n}\,e_k=0, \qq \&  \qq {\cal A}_0^{(n)}<1.\eeq
\end{nth}

\vspace{5mm}\nin{\bf Proof:} 
Corollary (4.16) in \cite{Be} ensures that the metric is globally defined on ${\mb S}^2$ and Zoll iff:
\begin{itemize}
\item ${\cal A}(\tht)$ is odd in terms of $x=\cos\tht\in[-1,+1]$,
\item ${\cal A}(0)={\cal A}(\pi)=0$.
\item ${\cal A}([0,\pi])\subset\,(-1,+1)$.
\end{itemize}
The first property is obvious and the second one follows from
\beq
{\cal A}(0)=-{\cal A}(\pi)=\sum_{k=1}^{2n}\,e_k=0.
\eeq
The third property follows from ${\cal A}_0^{(n)}<1$ and Lemma \ref{boundA}.

The hamiltonian is therefore globally defined as well as
$\Pi$ and $\Pf/s$. Let us write the integrals
\beq
S_1=\cos\phi\,{\cal S}+\sin\phi\,{\cal T}, \qq S_2=\cos\phi\,{\cal T}-\sin\phi\,{\cal S},\eeq 
where 
\beq
{\cal S}=\Pi\sum_{k=0}^n\,\la_{2k-1}(\tht)\,H^{n-k}\,\Pf^{2k},\quad \la_{-1}=1, \qq {\cal T}=\Pf\sum_{k=0}^n\,\la_{2k}(\tht)\,H^{n-k}\,\Pf^{2k}.
\eeq 
A look at Definition \ref{sollambda} shows that we can write
\beq
\la_{2k-1}=\frac{(-1)^k}{s^{2k}}\,\mu_{2k-1}  \qq \la_{2k}=\frac{(-1)^{k+1}}{s^{2k+1}}\,\mu_{2k},
\eeq
where the $\mu$'s are $C^{\nf}$ for $\tht\in\,[0,\pi]$. It follows that
\beq\left\{\barr{l}\dst 
{\cal S}=\sum_{k=0}^n\,(-1)^k\mu_{2k-1}(\tht)\,H^{n-k}\,
\Pi\,\left(\frac{\Pf}{s}\right)^{2k},\\[4mm]\dst   
{\cal T}=\sum_{k=0}^n(-1)^{k+1}\mu_{2k}(\tht)\,H^{n-k}\,\left(\frac{\Pf}{s}\right)^{2k+1},\earr\right.
\eeq 
are globally defined as well. $\hfill\Box$

\vspace{5mm}\nin{\bf Remarks}: 
\brm
\item Let us show that the set defined by the restriction $\ {\cal A}_0^{(n)}<1\ $ is not empty. Indeed the choice 
\[
\mu_{2k-1}=k^2+k-1,\qq \mu_{2k}=k^2(k+1)^2\qq\Longrightarrow\qq {\cal A}_0^{(n)}=\frac n{n+1}.\]
\item Let us give an example for which ${\cal A}_0^{(n)}>1$:  
\[
\mu_{2k-1}=1, \qq\qq \mu_{2k}=k^2(k+1)^2 \qq\Longrightarrow\qq {\cal A}_0^{(n)}=\frac{n^2}{n+1}>1 \quad \mbox{for}\quad n\geq 2.\] 
\item As stated in Theorem 1, the  constraint $|{\cal A}_0^{(n)}|<1$ is needed only for $n\geq 2$. Indeed for $n=1$ we can take
\beq
{\cal A}(\tht)=c\left(\frac 1{\sqrt{1-m_1\,s^2}}-\frac 1{\sqrt{1-m_2\,s^2}}\right),\qq\quad m_2<m_1<1.\eeq 
It is easy to prove that ${\cal A}([0,\pi])=[-f_0,+f_0]$ where
\beq
f_0=\frac{1-\rho}{\sqrt{(1-\rho)^2+3\rho}},\qq\qq 
\rho=\left(\frac{1-m_1}{1-m_2}\right)^{1/3}, \eeq
hence $f_0\,\in\,(-1,1)$.
\item In the proof of Lemma \ref{boundA} one can write alternatively
\[f_k(t)=\frac{(\sqrt{\mu_{2k}}-\sqrt{\mu_{2k-1}})}{\sqrt{\mu_{2k-1}}}
\,\frac 1{\sqrt{1+\mu_{2k}\,t^2}}\,\frac{\sqrt{\mu_{2k-1}}\,t}{\sqrt{1+\mu_{2k-1}\,t^2}}
\frac{\sqrt{\mu_{2k}}\,t+\sqrt{\mu_{2k-1}}\,t}{\sqrt{1+\mu_{2k}\,t^2}+\sqrt{1+\mu_{2k-1}\,t^2}},\]
which gives
\[
|f_k(t)|\leq\left|1-\sqrt{ \frac{\mu_{2k}}{\mu_{2k-1}} }\right|.  \]
So the bound 
\beq
{\cal B}_0^{(n)}\equiv\sum_{k=0}^{2n}\,\left|1-\sqrt{ \frac{\mu_{2k}}{\mu_{2k-1}} }\right|<1\eeq
does also ensure that $|A(\tht)|<1$ uniformly.
\erm

\subsection{The Poisson algebra}
Having defined
\beq
S_+=e^{-i\phi}({\cal S}+i{\cal T}), \qq\qq S_-=e^{i\phi}({\cal S}-i{\cal T}),\eeq
let us begin with:
\begin{ndef} The set of moments 
$\{\si_0,\si_1,\ldots,\si_{2n+1}\}$ and their generating function are defined by
\beq\label{defprod1}
S_+\,S_-\equiv{\cal S}^2+{\cal T}^2=\sum_{l=0}^{2n+1}\,\si_l\,H^{2n-l+1}\,\Pf^{2l},\qq \Si(\xi)=\sum_{l=0}^{2n+1}\,\si_l\,\xi^l, \qq \xi\in{\mb C}.
\eeq
\end{ndef}
The $\si_l$ are related to the $\la$'s by

\begin{nth}\label{siglambda} The moments are given by:
\beq\label{defsigma}\barr{cccl}
0 \leq l \leq n: & \qq \si_l & = & \dst 
\sum_{k=0}^{l}\,S_{k,l-k} \qq\quad(\Rightarrow\quad \si_0=1)\\[4mm]
n+1 \leq l \leq 2n+1: & \qq \si_l & = & \dst 
\sum_{k=l-n-1}^{n}\,S_{k,l-k},\earr\eeq
where
\beq\label{Skl}
S_{k,l}=\la_{2k-1}\,\la_{2l-1}+\la_{2k}\la_{2l-2}-\frac 1{s^2}\la_{2k-1}\la_{2l-3},
\eeq
and with the conventions that $\la_{2n+1}=\la_{-2}=\la_{-3}=0.$
\end{nth}

\vspace{5mm}\nin{\bf Proof:} Using the formulae given for the $S_1$ and $S_2$ (and taking into account the conventional values) we have
\beq\label{carres}
S_+\,S_-=\sum_{k=0}^n\sum_{L=0}^{n+1}\,S_{k,L}\,H^{2n-k-L+1}\Pf^{2(k+L)},
\eeq
where $S_{k,L}$ is given by (\ref{Skl}). 
The change of summation index $L \to l=L+k$ gives
\beq
S_+\,S_-=\sum_{k=0}^n\sum_{l=k}^{n+k+1}\,S_{k,l-k}\,H^{2n-l+1}\Pf^{2l}.
\eeq
Interchanging the order of the summations we get
\beq
S_+\,S_-=\sum_{l=0}^n\Big(\sum_{k=0}^l\,S_{k,l-k}\Big)H^{2n+1-l}\,\Pf^{2l}+\sum_{l=n+1}^{2n+1}\Big(\sum_{k=l-n-1}^n\,S_{k,l-k}\Big)H^{2n+1-l}\,\Pf^{2l},\eeq
from which the relations in (\ref{defsigma}) follow.
$\hfill \Box$

To relate the moments $\si_l$, hence their generating function $\Si(\xi)$, in terms of the parameters $m_k$ appearing in $A(\tht)$ several steps are needed. In the first one we need to relate $\Si(\xi)$ to the generating functions: 

\begin{nth}\label{Sigma} The generating function of the moments is given by  
\beq\label{formSi}
\Si(\xi)=\xi\,{\cal L}^2(\tht,\xi)+(1+\tau)\,{\cal M}^2(\tht,\xi), \qq\qq \tau=-\frac{\xi}{s^2}.
\eeq
\end{nth}

\vspace{5mm}
\nin{\bf Proof:} Using (\ref{defsigma}) we have
\beq
\Si(\xi)=\sum_{L=0}^n\,\xi^L\sum_{k=0}^{L}\,S_{k,L-k}+\sum_{L=n+1}^{2n+1}\xi^L\,\sum_{k=L-n-1}^n\,S_{k,L-k}.
\eeq
Interchanging the orders of the summations gives
\beq
\Si(\xi)=\sum_{k=0}^{n}\sum_{L=k}^{k+n+1}\,\xi^L\,S_{k,L-k}=\sum_{k=0}^{n}\,\xi^k\sum_{l=0}^{n+1}\,S_{k,l}\,\xi^l.
\eeq
The terms in $S_{k,l}$ give successively
\beq
\sum_{k=0}^n\xi^k\la_{2k-1}\sum_{l=0}^{n+1}\xi^l\la_{2l-1}={\cal M}^2, \qq 
\sum_{k=0}^n\xi^k\la_{2k}\sum_{l=0}^{n+1}\xi^l\la_{2(l-1)}=\xi\,{\cal L}^2,
\eeq
and
\beq
-\frac 1{s^2}\,\sum_{k=0}^n\xi^k\la_{2k-1}\sum_{l=1}^{n+1}\xi^l\la_{2l-3}=\tau{\cal M}^2.
\eeq
Adding all these pieces proves the Proposition. $\hfill\Box$

In a second step we need a new writing of the generating functions

\begin{nth} For $\tau\geq 0$ (that is for $\xi\leq 0$) one has for the first generating function
\beq\label{formL}
{\cal L}(\tht,\xi)=-\frac{(1+\tau)^n}{2\,\eta\,s}\Big((1+\eta\,c){\cal H}(\tht,\eta)-(1-\eta\,c){\cal H}(\tht,-\eta)\Big),\eeq
where
\[
\tau=-\frac{\xi}{s^2} ,\qq\quad 
\eta=\sqrt{\frac{\tau}{\tau+1}},\qq\quad {\cal H}(\tht,\xi)=\prod_{k=1}^{2n}(1+\xi\,h_k(\tht)).\]
The second generating function is given by
\beq\label{formM}
{\cal M}(\tht,\xi)=\frac{(1+\tau)^n}{2}\Big((1+\eta\,c){\cal H}(\tht,\eta)+(1-\eta\,c){\cal H}(\tht,-\eta)\Big).\eeq\end{nth}

\nin{\bf Proof:} The relation (\ref{gfL}), written out in detail gives
\beq
(-s){\cal L}=\sum_{l=0}^{n-1}\tau^l(1+\tau)^{n-l}(H)_{2l+1}+c\,\sum_{l=0}^n\tau^l(1+\tau)^{n-l}(H)_{2l},\eeq
which becomes
\beq
(-s){\cal L}=(1+\tau)^n\left(\frac 1{\eta}\sum_{l=0}^{n-1}\eta^{2l+1}(H)_{2l+1}+c\,\sum_{l=0}^n\eta^{2l}(H)_{2l}\right).\eeq
These sums are given by relations (\ref{evenH}) and (\ref{oddH}) in Appendix A, and lead to (\ref{formL}).
The proof of (\ref{formM}) is similar.$\hfill\Box$

Let us now express the moments in terms the parameters $m_k$ which appear in $A(\tht)$. 
To this end we will define, for the string $M=(m_1,m_2,\ldots,m_{2n})$ the symmetric functions 
$(M)_l$:
\beq
\prod_{l=1}^{2n}\,(1+\xi\,m_l)=\sum_{l=0}^{2n}\,\xi^l\,(M)_l.
\eeq
We are now in position to prove:

\begin{nth}\label{compsigma} The generating function of the moments is
\beq\label{Si1}
\Si(\xi)=(1-\xi)\,\prod_{l=1}^{2n}\,(1-\xi\,m_l),\qq\quad \xi\in{\mb C},
\eeq
giving the explicit formulae
\beq\label{tabSi1}\barr{cl}
      & \qq \si_0=1\\[4mm]
1\leq l\leq 2n: &\qq \si_l=(-1)^l[(M)_l+(M)_{l-1}]\\[4mm]              
      & \dst\qq\si_{2n+1}=-(M)_{2n}=
      -\prod_{k=1}^{2n}\,m_{k}.\earr
\eeq 
\end{nth}

\vspace{5mm}\nin{\bf Proof:} We will take $\xi\leq 0$ ensuring that $\tau\geq 0$. We have seen in (\ref{formSi}) that $\Si$ is given by
\beq
\Si(\xi)=(1+\tau){\cal M}^2-\tau(s{\cal L})^2.
\eeq
Upon use of relations (\ref{formL}) and (\ref{formM}) one obtains
\beq
\Si(\xi)=(1+\tau)^{2n+1}(1-\eta^2\,c^2)\,{\cal H}(\tht,\eta)\,{\cal H}(\tht,-\eta).\eeq
The identities
\beq
(1+\tau)(1-\eta^2\,c^2)=1-\xi, \qq (1+\tau)^{2n}\,{\cal H}(\tht,\eta)\,{\cal H}(\tht,-\eta)    =\prod_{k=1}^{2n}(1-\xi\,m_k),\eeq
lead for $\Si(\xi)$ to the relation (\ref{Si1}). Analytic continuation extends it to $\xi\in{\mb C}$. Expanding 
$\Si(\xi)$ in powers of $\xi$ gives (\ref{tabSi1}). $\hfill\Box$

\vspace{5mm}\nin{\bf Remark:} It follows that
\beq
\Si(1)=\sum_{l=0}^{2n+1}\,\si_l =0.\eeq

Let us conclude with:

\begin{nth}\label{comSS} One has the relation
\beq\label{commutS}
\{S_+,S_-\}=2i\,\sum_{l=0}^{2n}\,(l+1)\si_{l+1}\,H^{2n-l}\,\Pf^{2l+1}.\eeq
\end{nth}

\vspace{2mm}\nin{\bf Proof:} Extracting out from the bracket the $\phi$ dependence gives
\beq
\frac{\{S_+,S_-\}}{2i}=\frac 12\frac{\pt}{\pt \Pf}({\cal S}^2+{\cal T}^2)-\{{\cal S},{\cal T}\}.
\eeq
The first term in the right hand side gives
\beq
\sum_{l=0}^{2n}\,(l+1)\si_{l+1}\,H^{2n-l}\,\Pf^{2l+1}+\frac 1{s^2}\sum_{l=0}^{2n}(2n+1-l)\,\si_l\,H^{2n-l}\,\Pf^{2l+1},\eeq
so that the relation (\ref{commutS}) will hold true if we can prove the relation
\beq\label{null}
s^2\{{\cal S},{\cal T}\}=\sum_{l=0}^{2n}(2n+1-l)\,\si_l\,H^{2n-l}\,\Pf^{2l+1}.\eeq
Using the notation
$\Psi^{2n}_{k+l}=H^{2n-k-l}\,\Pf^{2(k+l)+1}$,
we have first
\beq\barr{l}\dst 
\{{\cal S},{\cal T}\}=\sum_{k,l}\Big[(n-k)\la_{2k-1}\Pi\{H,\la_{2l}\}-(n-l)\la_{2l}\{H,\la_{2k-1}\Pi\}+\\[4mm]\dst \hspace{8cm}+H\{\la_{2k-1}\Pi,\la_{2l}\}\Big]\frac{\Psi^{2n}_{k+l}}{H}.
\earr\eeq
In the second sum let us change $l\leftrightarrow k$, and let us notice that 
\beq
\{H,\la_{2l}\}=2\Pi\,\frac{\la'_{2l}}{A}=2\Pi\frac{\la_{2l-1}}{s^2},\eeq
thanks to relations (a) in (\ref{sd1}). Computing the other brackets gives
\beq\barr{l}\dst 
s^2\{{\cal S},{\cal T}\}=\sum_{k,l}\Big[2(n-k)\Pi^2\Big(\la_{2k-1}\la_{2l-1}-\la_{2k}\frac{s^2\la'_{2l-1}}{A}\Big)+\\[4mm]\dst 
\hspace{4cm}+2(n-k)\frac c{sA}\la_{2k}\la_{2l-1}\Pf^2+H\,\la_{2k-1}\la_{2l-1}\Big]\Psi^{2n}_{k+l}.\earr\eeq
Using $\dst \Pi^2=H-\frac{\Pf^2}{s^2}$ leads to
\beq\barr{l}\dst 
s^2\{{\cal S},{\cal T}\}=\sum_{k,l}\left[(2n-2k+1)\la_{2k-1}\la_{2l-1}-2(n-k)s^2\la_{2k}\frac{s^2\la'_{2l-1}}{A}\right]\Psi^{2n}_{k+l}+\\[4mm]\dst 
\hspace{1cm}+\sum_{k,l}\left[2(n-k)\frac{\la_{2k}}{A}\left(\la'_{2l-1}-\frac cs\la_{2l-1}\right)
-2(n-k)\frac{\la_{2k-1}\la_{2l-1}}{s^2}\right]
\Psi^{2n}_{k+l+1}.\earr
\eeq
Noticing that $\la'_{-1}=0$ we can write 
\beq
\sum_k\sum_{l=1}^n2(n-k)\la_{2k}\frac{s^2\la'_{2l-1}}{A}\Psi^{2n}_{k+l}=\sum_{k,l}2(n-k)\la_{2k}\frac{s^2\la'_{2l+1}}{A}\Psi^{2n}_{k+l+1}\eeq
since $\la_{2n+1}=0$. This allows to collect the three terms exhibiting a factor $1/A$ 
\beq
-\sum_{k,l}2(n-k)\frac{\la_{2k}}{A}\left(s^2\,\la'_{2l+1}-\la'_{2l-1}+\frac cs\,\la_{2l-1}\right)\Psi^{2n}_{k+l+1}\eeq
and upon use of relations (b) in (\ref{sd1}) these terms reduce to
\beq
\sum_{k,l}2(n-k)\la_{2k}\la_{2l}\,\Psi^{2n}_{k+l+1}.\eeq
So we end up with the left hand member of (\ref{null}):
\beq\label{piece1}\barr{l}\dst 
s^2\{{\cal S},{\cal T}\}=\sum_{k,l}(2n-2k+1)\la_{2k-1}
\la_{2l-1}\Psi^{2n}_{k+l}+\\[4mm]\dst
\hspace{5cm}+\sum_{k,l}2(n-k)\left(\la_{2k}\la_{2l}-\frac{\la_{2k-1}\la_{2l-1}}{s^2}\right)\Psi^{2n}_{k+l+1}.\earr\eeq
 
Let us consider now the right hand member of (\ref{null}) with $l \to L$. Exchanging the summation indices leads to 
\beq 
\sum_{L=0}^{2n+1}(2n-L+1)\si_L\,H^{2n-L}\,\Pf^{2L+1}=\sum_{k}\sum_{L=k}^{k+n+1}\,(2n-L+1)S_{k,L-k}\,H^{2n-L}\,\Pf^{2L+1}.
\eeq
The change of summation index: $l=L-k$ gives eventually
\beq
\sum_{k}\sum_{l=0}^{n+1}\,(2n-k-l+1)\,S_{k,l}\,\Psi^{2n}_{k+l}.\eeq
Let us recall that
\beq
S_{k,l}=\la_{2k-1}\la_{2l-1}+\la_{2k}\la_{2l-2}-\frac 1{s^2}\la_{2k-1}\la_{2l-3}.\eeq
The first term, due to the $k \leftrightarrow l$ 
symmetry, gives  
\beq
\sum_{k,l}\,(2n-2k+1)\la_{2k-1}\la_{2l-1}\,\Psi^{2n}_{k+l}\eeq
while the remaining two terms, with the change $l \to l+1$, give
\beq
\sum_{k,l}\,2(n-k)\left(\la_{2k}\la_{2l}-\frac 1{s^2}\la_{2k-1}\la_{2l-1}\right)\,\Psi^{2n}_{k+l+1}.\eeq
Adding these two terms and comparing with (\ref{piece1}) establishes the relation (\ref{null}), hence the Proposition. $\hfill\Box$

Combining relations (\ref{defprod1}) and (\ref{commutS}) we conclude to:

\begin{nth} The set of observables
\beq
H,\qq\quad \Pf, \qq\quad S_+= e^{-i\phi}\,({\cal S}+ i{\cal T}),\qq\quad S_-=e^{i\phi}\,({\cal S}\,-i{\cal T}),\eeq
is indeed a Poisson algebra with
\beq
S_+S_-=\sum_{l=0}^{2n+1}\,\si_l\,H^{2n-l+1}\,\Pf^{2l}, \qq \{S_+,S_-\}=2i\,\sum_{l=0}^{2n}\,(l+1)\si_{l+1}\,H^{2n-l}\,\Pf^{2l+1}.
\eeq
\end{nth}

\vspace{5mm}
\nin{\bf\large This concludes the proof of Theorem 1}.$\hfill\Box$

\nc{\Iph}{I_{\phi}}  \nc{\Ith}{I_{\tht}}
\section{Geodesics}
\subsection{Geodesics from the action-angle coordinates}
Since the hamiltonian defined by (\ref{metGV}) and (\ref{metGV1}) is globally defined on $M={\mb S}^2$ it is interesting to study its geodesics

Before computing the action and angle variables let us consider the torus $H=E$ and $\Pf=L$. The Hamilton-Jacobi equation for the action:
\beq
\frac{\pt\,S}{\pt\,t}+\frac 1{A^2}\,\left(\frac{\pt\,S}{\pt\,\tht}\right)^2+\frac 1{s^2}\left(\frac{\pt\,S}{\pt\,\phi}\right)^2=0,\eeq
allows for separation
\beq
S=-E\,t+L\,\phi+\int\,\Pt\,d\tht\eeq
and leads to 
\beq
\Pt=\eps\sqrt{E}\sqrt{1-\frac{s_0^2}{s^2}}(1+{\cal A}(\tht)),\qq s_0=\sin i=\frac L{\sqrt{E}}, \qq i\in\,(0,\frac{\pi}{2}).\eeq
The sign $\eps$ is given by
\beq
\eps=\left\{\barr{lcl} +1 & \quad\mbox{if}\quad & \quad \tht:\ i\ \to \pi-i\\[4mm] -1 & \quad\mbox{if}\quad & \quad \tht:\ \pi-i\ \to\ i.\earr\right.\eeq
As may be seen in Figure 1, the plus sign corresponds to the first half of the geodesic where $\tht$ increases  from $i$ to $\pi-i$, while the minus sign corresponds to the second half of the geodesic where $\tht$ decreases from $\pi-i$ to $i$.

\begin{figure}[H]
\begin{center}
\includegraphics[scale=0.5]{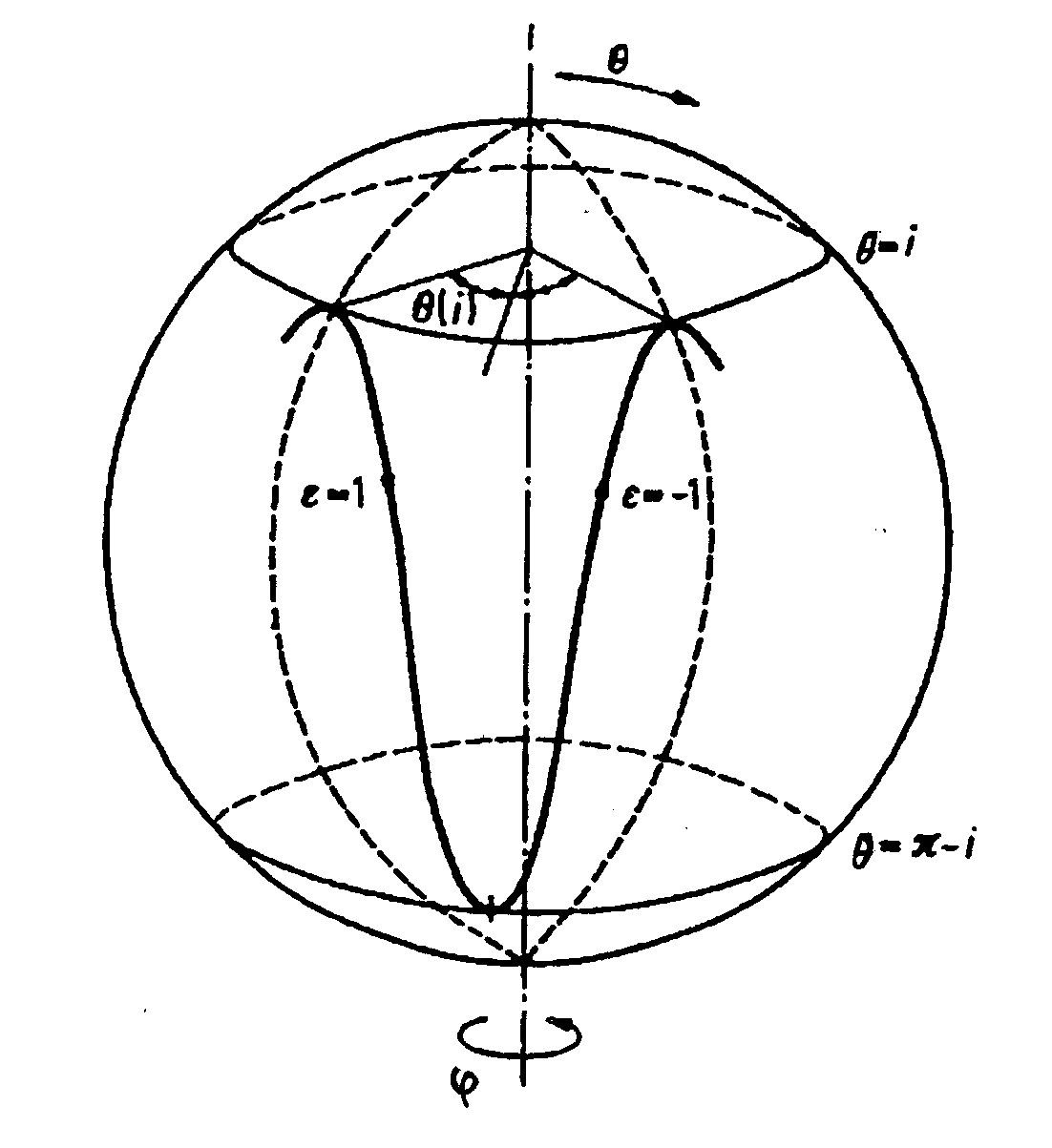}
\end{center}
\caption{Geometry of the geodesics}
\end{figure}

We will take the initial values:
\[   \tht=i\qq s=s_0 \qq c=c_0\qq \phi=0.   \]  
The first action is
\beq
\Iph=\frac 1{2\pi}\oint \,L\,d\phi=L,\eeq
and the second one
\beq
\Ith=\frac 1{2\pi}\,\oint\,\Pt\,d\tht=\frac{\sqrt{E}}{\pi}\int_i^{\pi-i}\,\sqrt{1-\frac{s_0^2}{s^2}}\,d\tht
\eeq
because ${\cal A}(\tht)$ is odd.
 coordinate change $\dst\sin\chi=\frac c{c_0}$ gives
\beq
\Ith=2\sqrt{E}\,\frac{c_0^2}{\pi}\int_0^{\pi/2}\frac{\cos^2\chi}{1-c_0^2\,\sin^2\chi}\,d\chi=\sqrt{E}-L.\eeq
Hence we have obtained for the actions
\beq
\left\{\barr{l} \Iph=\Pf \\[4mm] \Ith=\sqrt{H}-\Pf\earr \right.\qq\Longrightarrow\qq H=(\Ith+\Iph)^2.
\eeq
Due to the superintegrability, the dynamical system is degenerate and we have a single frequency
\beq
\nu=\frac{\pt H}{\pt \Ith}=2(\Ith+\Iph)=2\sqrt{E}\qq\Longrightarrow\qq \om_{\tht}=\nu\,t+K,\qq \om_{\phi}=\nu\,t+L,
\eeq
which determines the time dependence of the angles.

\vspace{5mm}\nin {\bf Remark}: Since ${\cal A}(\tht)$ is odd, it does not contribute to the action integrals.  Hence the previous relations for the actions are in fact valid for {\em any Zoll metric of revolution}.

\nc{\omt}{\om_{\tht}}  \nc{\omp}{\om_{\phi}}

Let us now determine the angles:
\begin{nth} For the first half of the geodesic one has
\beq\label{omtheta}
2\sqrt{E}\,t=\arccos\left(\frac c{c_0}\right)+\sum_{k=1}^{2n}\,e_k\,\Om(\tht,m_k),\eeq
where
\beq
\Om(\tht,m_k)=\left\{
\barr{ccl}\dst \frac 1{\sqrt{m_k}}\,\arcsin\left(\sqrt{\frac{m_k}{1-m_k\,s_0^2}}\sqrt{s^2-s_0^2}
\right) \quad & \quad \mbox{if} \quad & m_k\in(0,1)
\\[5mm]\dst \sqrt{s^2-s_0^2} \quad & \quad \mbox{if} \quad & m_k=0\\[5mm]\dst \frac 1{\sqrt{|m_k|}}\,{\rm arcsinh}\left(\sqrt{\frac{|m_k|}{1+|m_k|\,s_0^2}}
\sqrt{s^2-s_0^2}\right) \quad & \quad \mbox{if} \quad & m_k<0.\earr\right.\eeq
\end{nth}

\nin{\bf Proof:} We have, for the first half of the geodesic
\beq
\om_{\tht}=\frac{\pt S}{\pt \Ith}=\int\,\frac{(1+{\cal A}(\tht))}{\sqrt{s^2-s_0^2}}\,s\,d\tht,\eeq
so that
\beq
\om_{\tht}=\int\frac s{\sqrt{s^2-s_0^2}}d\tht+\sum_{k=1}^{2n}\,e_k\int\frac{sc}{\sqrt{s^2-s_0^2}\,\sqrt{1-m_k\,s^2}}\,d\tht.
\eeq
The second integral requires the change of variable $u=\sqrt{s^2-s_0^2}$.$\hfill\Box$

From this we deduce
\begin{nth}[Kepler's law] The period of the geodesic motion is given by $\ \dst T=\frac{\pi}{\sqrt{E}}$.
\end{nth}

\nin{\bf Proof:} When $\tht$ increases from $i$ to 
$\pi-i$ the time evolves from $t=0$ to $\dst t=\frac T2$ 
while the right hand member in (\ref{omtheta}) evolves from $0$ to $\pi$. $\hfill\Box$

Let us compute now the angle $\om_{\phi}$ which is more 
interesting since it will give a first description of the geodesics:
\begin{nth} The analytic structure of the geodesics, when $\tht$ increases from $i$ to $\pi-i$ and $\phi$ from $0$ to $\pi$, is given by:
\beq\label{geoinc}
\phi=\arccos\left(\frac{s_0}{s}\frac c{c_0}\right)+\sum_{k=1}^{2n}\,e_k\,\arcsin\left(\frac 1{\sqrt{1-m_k\,s_0^2}}\sqrt{1-\frac{s_0^2}{s^2}}\right),\eeq
and when $\tht$ decreases from $\pi-i$ to $i$ while 
$\phi$ increases from $\pi$ to $2\pi$, is given by:
\beq\label{geodec}
\phi=2\pi-\arccos\left(\frac{s_0}{s}\frac c{c_0}\right)-\sum_{k=1}^{2n}\,e_k\,\arcsin\left(\frac 1{\sqrt{1-m_k\,s_0^2}}\sqrt{1-\frac{s_0^2}{s^2}}\right).\eeq
It follows that all the geodesics are closed.
\end{nth}

\nin{\bf Proof:} We have
\beq
\om_{\phi}=\frac{\pt S}{\pt \Iph}=\phi+\omt-s_0\,\eps\,\int\,\frac{(1+{\cal A}(\tht))}{s\,\sqrt{s^2-s_0^2}}\,d\tht.\eeq
The change of variable $\dst u=\frac{\sqrt{s^2-s_0^2}}{s}$ allows to evaluate the integral and gives
\beq
\phi=L-K+\eps\,\arccos\left(\frac{s_0}{s}\frac c{c_0}\right)+\eps\sum_{k=1}^{2n}\,e_k\,\arcsin\left(\frac 1{\sqrt{1-m_k\,s_0^2}}\sqrt{1-\frac{s_0^2}{s^2}}\right).\eeq
When $\tht$ increases from $i$ (starting with $\phi=0$) to $\pi-i$ we will have $\eps=+1$, hence (\ref{geoinc}).
When $\tht$ decreases from $\pi-i$ (starting with $\phi=\pi$) to $i$ we have $\eps=-1$, hence (\ref{geodec}). At the end of the turn $\phi$ has increased from $0$ to $2\pi$ and the geodesic does close, as it should, since  the metric is Zoll.$\hfill\Box$

For future use let us prove

\begin{nth} When $\tht$ increases from $i$ to $\pi-i$ one has
\beq\label{geod1}
K\,e^{i\phi}=\left(s_0\frac cs+i\sqrt{1-\frac{s_0^2}{s^2}}\right)\prod_{k=1}^{2n}\left(\frac{s_0}{s}\sqrt{1-m_k\,s^2}+ie_k\sqrt{1-\frac{s_0^2}{s^2}}\right),
\eeq
where
\[ K=c_0\prod_{k=1}^{2n}\,\sqrt{1-m_k\,s_0^2}.\]
\end{nth}

\nin{\bf Proof:} This exponential produces as a first term
\beq
\exp{\dst \left(i\,\arccos(\frac{s_0\,c}{c_0\,s})\right)}=\frac 1{c_0}\left(s_0\frac cs+i\sqrt{1-\frac{s_0^2}{s^2}}\right)
\eeq
multiplied by the product involving the factors
\beq
\exp{\left(i\arcsin\,(\frac 1{\sqrt{1-m_k\,s_0^2}}\sqrt{1-\frac{s_0^2}{s^2}})\right)}=\frac 1{\sqrt{1-m_k\,s_0^2}}\left(\frac{s_0}{s}\sqrt{1-m_k\,s^2}+ie_k\sqrt{1-\frac{s_0^2}{s^2}}\right),\eeq
leading to (\ref{geod1}).$\hfill\Box$

\vspace{3mm}\nin{\bf Remark:} this gives another description of the first half of the geodesics. For the second half it is sufficient to change $\phi\ \to\ 2\pi-\phi$ in (\ref{geod1}).

\subsection{Geodesics from the integrals} 
As pointed out in \cite{Va2} for the cubic case, one can recover rather conveniently the geodesics from the very extra integrals. It is therefore interesting to check that, by this rather different approach, we do get the relation (\ref{geod1}) for the first half of the geodesic.

On the torus $H=E$ and $\Pf=L$ we have
\beq
\frac{S_1+iS_2}{E^{n+1/2}}=e^{-i\phi}\frac{({\cal S}+i{\cal T})}{E^{n+1/2}}.
\eeq
These quantities are easily extracted out from the generating functions 
\beq
\frac{{\cal T}}{E^{n+1/2}}=\frac L{\sqrt{E}}\sum_{k=0}^n\,\la_{2k}(\tht)\,\left(\frac{L^2}{E}\right)^k=s_0\,{\cal L}(\tht,s_0^2),\eeq
where ${\cal L}$ is given by (\ref{gfL}), and 
\beq
\frac{{\cal S}}{E^{n+1/2}}=\sqrt{1-\frac{s_0^2}{s^2}}\sum_{k=0}^n\,\la_{2k-1}(\tht)\,\left(\frac{L^2}{E}\right)^k=\sqrt{1-\frac{s_0^2}{s^2}}\,{\cal M}(\tht,s_0^2),\eeq
where ${\cal M}$ is given by (\ref{gfM}). One can write
\beq\barr{l}\dst 
(-s){\cal L}(\tht,s_0^2)=\left(\sqrt{1-\frac{s_0^2}{s^2}}\right)^{2n}\left(\sum_{k=0}^{n-1}(-1)^k\mu^{2k}\,(H)_{2k}(\tht)+\right.\\[4mm]\dst 
\hspace{7cm}\left.+\frac c{\mu}\sum_{k=0}^n\,(-1)^k\,\mu^{2k+1}(H)_{2k+1}(\tht)\right),\earr
\eeq
with $\dst \mu=\frac{s_0}{\sqrt{s^2-s_0^2}}.$ Using the relations (\ref{evenH}) and (\ref{oddH}) with $\xi\ \to\ i\mu$ we obtain
\beq\label{geoL}
is_0\,{\cal L}(\tht,s_0^2)=-\sqrt{1-\frac{s_0^2}{s^2}}\,\frac 12({\cal P}-\ol{\cal P})+s_0\frac cs\,\frac 1{2i}({\cal P}+\ol{\cal P}),
\eeq
where
\[  {\cal P}=\prod_{k=1}^{2n}\left(\sqrt{1-\frac{s_0^2}{s^2}}+i\,e_k\,\frac{s_0}{s}\,\sqrt{1-m_k\,s^2}\right).\]
It remains to compute
\beq\label{geoM}\barr{l}\dst 
{\cal M}(\tht,s_0^2)=\left(\sqrt{1-\frac{s_0^2}{s^2}}\right)^{2n}\left(\sum_{k=0}^{n-1}(-1)^k\mu^{2k}\,(H)_{2k}(\tht)\right.\\[4mm]\dst 
\hspace{7cm}\left.-c\,\mu\sum_{k=0}^{n-1}\,(-1)^k\,\mu^{2k+1}\,(H)_{2k+1}(\tht)\right),\earr
\eeq
which leads to
\beq
\sqrt{1-\frac{s_0^2}{s^2}}\,{\cal M}(\tht,s_0^2)=\sqrt{1-\frac{s_0^2}{s^2}}\ \frac 12({\cal P}+\ol{\cal P})-s_0\frac cs\,\frac 1{2i}\,({\cal P}-\ol{\cal P}).\eeq
Hence we conclude to
\beq
\frac{S_1+iS_2}{E^{n+1/2}}=e^{-i\phi}\left[\sqrt{1-\frac{s_0^2}{s^2}}-is_0\frac cs\right]\,{\ol{\cal P}},
\eeq
which can be written
\beq
(-i)(-1)^n\,e^{-i\phi}\left[s_0\frac cs+i\,\sqrt{1-\frac{s_0^2}{s^2}}\right]\prod_{k=1}^{2n}\left(\frac{s_0}{s}\sqrt{1-m_k\,s^2}+ie_k\,\sqrt{1-\frac{s_0^2}{s^2}}\right).
\eeq
This conserved quantity, evaluated for $t=0$ and $s=s_0$, has for value
\beq 
(-i)(-1)^nc_0\prod_{k=1}^{2n}\sqrt{1-m_k\,s_0^2}=(-i)(-1)^n\,K.\eeq
And we do recover the relation (\ref{geod1}).$\hfill\Box$ 

\vspace{5mm}
\nin Let us proceed to the second part of this article,  devoted to the proof of Theorem 2.
\section{Integrals of even degree in the momenta}
Before dealing with the general case, let us first consider integrals which are quadratic in the momenta i. e. one of the Koenigs SI models \cite{Ko}. Using the coordinates of \cite{Va1} one has 
\beq
H^{(K)}=\frac{\cosh^2 x}{\rho+\sinh x}(P_x^2+P_y^2)\qq\quad \rho>0,
\eeq
and we will consider only the first extra integral
\beq\label{easy}
S_1^{(K)}=\cos y(H^{(K)}-2\sinh x\,P_y^2)+2\sin y\,\cosh x\,P_x\,P_y.
\eeq
In the coordinates used throughout this work, and anticipating on the results of the next sections, we have
\beq
H=\Pi^2+\frac{\Pf^2}{\sin^2\tht}, \qq \Pi=\frac{\Pt}{A(\tht)}, \qq A(\tht)=1+{\cal A}(\tht), \qq {\cal A}=\frac c{\sqrt{1-m\,s^2}},
\eeq
and for the first extra integral
\beq\label{intK}
S_1=\cos\phi\left(H-\frac{(1+c\sqrt{1-m\,s^2})}{s^2}\,\Pf^2\right)-\sin\phi\,\frac{(c+\sqrt{1-m\,s^2})}{s}\,\Pi\,\Pf, \quad m<1.\eeq
That this metric is not globally defined on ${\mb S}^2$ stems from the fact that ${\cal A}([0,\pi])=[-1,+1]$ instead of ${\cal A}([0,\pi])\subset\,(-1,+1)$. Nevertheless, these two metrics are related by the following local diffeomorphism:

\begin{nth} Provided that $m<0$ one has 
\[H^{(K)}=H/\la^2,\qq\quad 2\la^2=\sqrt{|m|}=\rho+\sqrt{\rho^2+1},\]
and
\beq
e^x=\frac 1{\sqrt{|m|}}\,\frac{1+\sqrt{1+|m|\,s^2}}{1-c},\qq \la^2\,s^2=\frac{\rho+\sinh x}{\cosh^2 x}.\eeq
\end{nth}

\vspace{5mm}\nin{\bf Proof:} Elementary computational check.$\hfill\Box$

\vspace{5mm}\nin{\bf Remarks:}
\brm
\item The coordinates $(\tht,\phi)$ appear rather weird when compared to the coordinates $(x,y)$ which lead to a simple structure for the integrals given in (\ref{easy}).
\item The fact that for trigonometric integrals the metric is not globally defined on ${\mb S}^2$ was first observed in \cite{Va1}. However, in this same reference, it was shown that there could be, for special choices of the parameters of Koenigs models, SI systems globally defined either on 
${\mb R}^2$ or on ${\mb H}^2$ which {\em cannot be obtained in our approach} since, as shown in \cite{Be},   the metric structure
\beq
g=A^2\,d\tht^2+\sin^2\tht\,d\phi^2 \qq\qq \tht\in\,(0,\pi) \qq \phi\in\,{\mb S}^1,
\eeq
is locally adapted only to ${\mb S}^2$.
\erm

Let us turn ourselves to the general case where 
$\sharp(S_1)=\sharp(S_2)=2n$ for $n\geq 1$. The hamiltonian remains
\beq\label{ham2}
H=\frac 12\left(\Pi^2+\frac{\Pf^2}{\sin^2\tht}\right)\qq\qq \Pi=\frac{\Pt}{A(\tht)}.
\eeq
The extra integrals will be again
\beq
S_1=\cos\phi\,{\cal S}+\sin\phi\,{\cal T}, \qq S_2=\cos\phi\,{\cal T}-\sin\phi\,{\cal S},\eeq
but this time we have
\beq
{\cal S}=\sum_{k=0}^n\,\la_{2k-1}(\tht)\,H^{n-k}\,\Pf^{2k},\qq\qq {\cal T}=\Pi\,\sum_{k=0}^{n-1}\,\la_{2k}(\tht)\,H^{n-k-1}\,\Pf^{2k+1},
\eeq
defining an array of functions of $\tht$ of the form
\[\left(\barr{ccccc}
\la_{-1}=1 &\quad \la_1 & \quad \la_3 &\quad \ldots & \la_{2n-1}\\[4mm]
 & \quad \la_0 & \quad \la_2 & \quad\ldots & \quad \la_{2(n-1)}\earr\right).\]

\nin{\bf Remark:} Since most of the proofs are similar to those of Section 3, we will proceed speedily.

\begin{nth} $S_1$ and $S_2$ will be integrals iff the 
$\la$'s s solve the differential system:
\beq\label{sd2}
0\leq k \leq n: \qq\left\{\barr{ccll}
s^2\,\la'_{2k-1} & = & -A\,\la_{2(k-1)} & \qq (a)
\\[4mm]
s^2\,\la'_{2k} & = & \dst\la'_{2(k-1)}-\frac cs\,\la_{2(k-1)}+A\,\la_{2k-1}, & \qq (b)
\earr\right.
\eeq
with the conventional values $\la_{2n}=\la_{-2}=0$.
\end{nth}

\vspace{5mm}
\nin{\bf Proof}: Similar to the proof of Proposition \ref{sd11} in Section 3. $\hfill\Box$

\vspace{5mm}\nin{\bf Remark:} 
Here too one can get rid of the derivatives in the right hand side of relation (\ref{sd2}) by a simple recurrence which gives for $0\leq k\leq n-1$:
\beq
s^{2(k+1)}\,\la'_{2k}=-sc(\la_0+s^2\la_2+\ldots+s^{2(k-1)}\la_{2(k-1)})+(1+s^2\la_1+\ldots+s^{2k}\la_{2k-1})A,
\eeq
while for $k=n$ one gets the purely algebraic relation
\beq\label{alg2}
sc(\la_0+s^2\la_2+\ldots+s^{2(n-1)}\la_{2(n-1)})= (1+s^2\la_1+\ldots+s^{2n}\la_{2n-1})A.
\eeq 

Let us consider again the generating functions, defined this time by:
\begin{ndef} The generating functions are now
\beq\label{defgf2}
{\cal L}(\tht,\xi)=\sum_{k=0}^{n-1}\la_{2k}(\tht)\,\xi^k \qq\qq {\cal M}(\tht,\xi)=\sum_{k=0}^n\la_{2k-1}(\tht)\,\xi^k.
\eeq\end{ndef}

\begin{nth}\label{sdiff2} The differential system (\ref{sd2}) is equivalent, in terms of generating functions, to
\beq
(1+\tau)\pt_{\tht}{\cal L} -\tau\frac cs{\cal L}=\frac A{s^2}\,{\cal M}, \qq\qq \pt_{\tht}{\cal M}=\tau\,A\,{\cal L},\qq\qq \tau=-\frac{\xi}{s^2}.\eeq
\end{nth}

\vspace{5mm}
\nin{\bf Proof:} Similar to the proof of Proposition \ref{odegf}. $\hfill\Box$

\subsection{The solution for integrals of even degree}
For $n\geq 1$, let us first define the functions
\beq\label{hk2}
\forall k\in\{1,2,\ldots,2n-1\}: \qq h_k(\tht)=e_k\,\sqrt{1-m_k\,s^2},\qq m_k<1,\eeq
and
\beq\label{gfH2}
{\cal H}(\tht,\xi)\equiv \prod_{k=1}^{2n-1}(1+\xi\,h_k(\tht))=\sum_{k=0}^{2n-1}\,(H)_k(\tht)\,\xi^k.
\eeq
The $(H)_k$ are nothing but the symmetric functions constructed in terms of the $h_k$. In terms of these objects, we can write down the solution for the $\la$'s. 

\begin{ndef}\label{sollambda2} Let us define, for $k\in\{0,1,2,\ldots,n-1\}$ 
\beq\label{lapair2}
\la_{2k}=\frac{(-1)^{k+1}}{s^{2k+1}}\left\{\sum_{l=0}^k\,{n-l-1 \choose n-k-1}\,(H)_{2l+1}+c\,\sum_{l=0}^{k}\,{n-l-1  \choose n-k-1}\,(H)_{2l}\right\},
\eeq
for $k\in\,\{0,1,\ldots,n-2\}$ \footnote{This set is empty for $n=1$.}
\beq\label{laimpair2}
\la_{2k+1}=\frac{(-1)^{k+1}}{s^{2(k+1)}}\left\{\sum_{l=0}^{k+1}\,{n-l \choose n-k-1}\,(H)_{2l}+c\sum_{l=0}^k\,{n-l-1 \choose n-k-1}\,(H)_{2l+1}\right\}
\eeq
and for $k=n-1$ 
\beq
\la_{2n-1}=\frac{(-1)^{n}}{s^{2n}}\left\{\sum_{l=0}^{n-1}\,(H)_{2l}+c\sum_{l=0}^{n-1}\,(H)_{2l+1}\right\}.\eeq
\end{ndef}

\vspace{3mm}{\bf Remark:} For $n=1$ we have a single parameter $m_1=m$ and the previous formulae give
\beq
\la_0=-\frac 1s((H)_1+c) \qq \la_1=-\frac 1{s^2}(1+c(H)_1)\qq (H)_1=\sqrt{1-m\,s^2}.
\eeq
Since the integrals are given by
\beq
{\cal S}=H^2+\la_1\,\Pf^2\qq\qq{\cal T}=\la_0\,\Pi\,\Pf,
\eeq
we have proved the relation anticipated in (\ref{intK}). 

Let us compute the generating functions.

\begin{nth} The generating functions ${\cal L}$ and ${\cal M}$ are given by
\beq\label{gfL2}
(-s){\cal L}(\tht,\xi)=\sum_{l=0}^{n-1}\,\psi_{l,n-1}\,(H)_{2l+1}+c\,\sum_{l=0}^{n-1}\,\psi_{l,n-1}\,(H)_{2l}
\eeq
and by
\beq\label{gfM2}
{\cal M}(\tht,\xi)=\sum_{l=0}^{n-1}\,\psi_{l,n}\,(H)_{2l}+c\,\sum_{l=0}^{n-1}\,\psi_{l+1,n}\,(H)_{2l+1}
\eeq
where
\beq
\tau=-\frac{\xi}{s^2}, \qq\qq \psi_{l,n}=\tau^l(1+\tau)^{n-l}, \qq  0\leq l \leq n.
\eeq
\end{nth}

\vspace{5mm}
\nin{\bf Proof:} From the definition of ${\cal L}$ and upon use of the formulae in (\ref{lapair2}) we have, after an exchange of the summation orders
\beq
(-s){\cal L}=\sum_{l=0}^{n-1}(H)_{2l+1}\sum_{k=l}^{n-1}{n-l-1 \choose n-k-1}\tau^k+c\sum_{l=0}^{n-1}(H)_{2l}\sum_{k=l}^{n-1}{n-l-1 \choose n-k-1}\tau^k .
\eeq 
Noticing that
\beq
\sum_{k=l}^{n-1}{n-l-1 \choose n-k-1}\tau^k=\tau^l\sum_{s=0}^{n-l-1}{n-l-1 \choose n-l-1-s}\tau^s=\tau^l(1+\tau)^{n-l-1}=\psi_{l,n-1}, 
\eeq 
implies relation (\ref{gfL2}).

In ${\cal M}$ the piece having $c$ factored out is
\beq
\sum_{k=1}^{n}\tau^k\sum _{l=0}^{k-1}{n-l-1 \choose n-k}(H)_{2l+1}=\sum_{l=0}^{n-1}\sum_{k=l+1}^n{n-l-1 \choose n-k}\tau^k=\sum_{l=0}^{n-1}\,\psi_{l+1,n}\,(H)_{2l+1}.
\eeq 
The remaining piece is
\beq 
\sum_{k=0}^{n-1}\tau^k\sum_{l=0}^k{n-l \choose n-k}\,(H)_{2l} +\tau^n\sum_{l=0}^{n-1}(H)_{2l}=\sum_{l=0}^{n-1}(H)_{2l}\sum_{k=l}^{n}{n-l \choose n-k}\tau^k.
\eeq 
The last term, using again the binomial theorem, becomes
\beq
\sum_{l=0}^{n}\,\psi_{l,n}\,(H)_{2l},
\eeq
and this ends up the proof.$\hfill\Box$ 
 
\vspace{3mm}{\bf Remark:} Here too, using these generating functions, one can check easily the algebraic relation (\ref{alg2}) which is
\beq
\left(sc\,{\cal L}-A\,{\cal M}\right)\Big|_{\tau=-1}=0.
\eeq 
 
Now let us use Proposition \ref{sdiff2} to prove

\begin{nth} The generating functions ${\cal L}$ and 
${\cal M}$, given by (\ref{gfL2}) and (\ref{gfM2}), are solutions of the equations
\beq\label{eq2et32} 
(1+\tau)\,\pt_{\tht}{\cal L}-\tau\frac cs\,{\cal L}=\frac A{s^2}\,{\cal M},\qq\qq \pt_{\tht}{\cal M}=\tau\,A\,
{\cal L},  \eeq
where
\beq
A(\tht)=1+c\sum_{k=1}^{2n-1}
\frac{e_k}{\sqrt{1-m_k\,s^2}}.\eeq
It follows that the $\la$'s given by  Definition \ref{sollambda2} are indeed a solution of the differential system (\ref{sdiff2}), and this implies that $S_1$ and $S_2$ are indeed integrals for the hamiltonian (\ref{ham2}).
\end{nth} 

\vspace{5mm}
\nin{\bf Proof:} Similar to the proof of Proposition \ref{GF1} in Section 3. One has to define the splitting
\beq
{\cal L}_1=-\frac 1s\sum_{l=0}^{n-1}\,\psi_{l,n-1}\,(H)_{2l+1}\qq\qq {\cal L}_2=-\frac cs\sum_{l=0}^{n-1}\,\psi_{l,n-1}\,(H)_{2l} \eeq
and similarly
\beq
{\cal M}_1=\sum_{l=0}^{n-1}\,\psi_{l,n}\,(H)_{2l}\qq\qq {\cal M}_2=c\,\sum_{l=0}^{n-1}\,\psi_{l+1,n}\,(H)_{2l+1}.\eeq
Let us begin with the derivative of ${\cal L}_1$. We have
\beq
-(1+\tau)\pt_{\tht}(s^{-1})\,\Psi_{l,n-1}=\frac c{s^2}\,\Psi_{l,n},\eeq
and
\beq 
(1+\tau)\pt_{\tht}\Psi_{l,n-1}=-\frac cs\,2\tau(1+\tau)\pt_{\tau}\Psi_{l,n-1}=-\frac cs\Big[2l\Psi_{l,n}+2(n-l-1)\Psi_{l+1,n}\Big],
\eeq
while the derivatives of $(H)_{2l+1}$ are given in Appendix A with $\nu=2n-1$. Adding all the terms we get
\beq
(1+\tau)\pt_{\tht}{\cal L}_1=\frac{(A-1)}{s^2}.\eeq
Observing that
\beq
\tau\psi_{l,n-1}=\psi_{l+1,n}\qq\Rightarrow\qq -\tau\frac cs{\cal L}_1=\frac{{\cal M}_2}{s^2},\eeq
we obtain 
\beq
(1+\tau)\pt_{\tht}{\cal L}_1-\tau\frac cs\,{\cal L}_1=\frac{{\cal M}_2}{s^2}+\frac{(A-1)}{s^2}\,{\cal M}_1.
\eeq
Similarly one can prove
\beq
(1+\tau)\pt_{\tht}{\cal L}_2-\tau\frac cs\,{\cal L}_2=\frac{{\cal M}_1}{s^2}+\frac{(A-1)}{s^2}\,{\cal M}_2.\eeq
The sum of these two relations proves (\ref{eq2et32}).

Let us proceed with the derivative of ${\cal M}_1$. We have
\beq
\pt_{\tht}\Psi_{l,n}=-\frac cs\Big[2l\Psi_{l,n}+2(n-l)\Psi_{l+1,n}\Big]\eeq
and using Appendix A we get, after easy algebra and use of (\ref{idH}):
\beq 
\pt_{\tht}{\cal M}_1=-\frac cs\sum_{l=0}^{n-1}
\Psi_{l+1,n}(H)_{2l}-\frac{(A-1)}{s}\sum_{l=0}^{n-1}\psi_{l+1,n}(H)_{2l+1}=\tau{\cal L}_2+\tau(A-1){\cal L}_1.
\eeq
Similarly one can prove

\beq
\pt_{\tht}{\cal M}_2=\tau\,{\cal L}_1+\tau(A-1){\cal L}_2.\eeq 
From which the second relation in (\ref{eq2et32}) follows. $\hfill\Box$

\subsection{The Poisson algebra}
Let us begin with:
\begin{nth} One defines the moments $\si_l$ and their generating functions as follows
\beq\label{Si2}
S_+\,S_-={\cal S}^2+{\cal T}^2=\sum_{l=0}^{2n}\si_l\,H^{2n-l}\,\Pf^{2l},\qq \Si(\xi)=\sum_{l=0}^{2n}\si_l\,\xi^l.
\eeq
These moments are related with the $\la's$ according to
\beq\label{defsigma2}\barr{cccl}
l=0: & \qq \si_0 & = & 1 \\[4mm]
1 \leq l \leq n: & \qq \si_l & = & \dst 
\sum_{k=0}^{l}\,S_{k,l-k} \\[4mm]
n+1 \leq l \leq 2n: & \qq \si_l & = & \dst 
\sum_{k=l-n}^{n}\,S_{k,l-k},\earr\eeq
where
\beq\label{Skl2}
S_{k,l}=\la_{2k-1}\,\la_{2l-1}+\la_{2(k-1)}\la_{2l}-\frac 1{s^2}\la_{2(k-1)}\la_{2(l-1)},
\eeq
and with the convention that $\la_{-2}=0$.
\end{nth}

\nin{\bf Proof:} Since $\la_{-2}=0$ it is convenient to write
\beq
{\cal T}=\Pi\sum_{k=0}^{n-1}\la_{2k}H^{n-k-1}\Pf^{2k+1}=\Pi\sum_{k=0}^n\la_{2(k-1)}H^{n-k}\Pf^{2k-1}.\eeq
It follows that
\beq
S_+S_-=\sum_{k,L=0}^nS_{k,L}H^{2n-k-L}\Pf^{2(k+L)},\eeq
where $S_{k,l}$ is given by (\ref{Skl2}). Defining 
$l=L+k$ we get
\beq
S_+S_-=\sum_{k=0}^n\sum_{l=k}^{n+k}S_{k,l-k}H^{2n-l}\Pf^{2l},\eeq
and upon exchange of the summations we end up with
\beq
S_+S_-=\sum_{l=0}^n \left(\sum_{k=0}^l S_{k,l-k}\right)H^{2n-l}\Pf^{2l}+\sum_{l=n+1}^{2n}\left(\sum_{k=l-n}^nS_{k,l-k}\right)H^{2n-l}\Pf^{2l},\eeq
which concludes the proof. $\hfill\Box$
    
Let us compute $\Si(\xi)$ in terms of the generating functions.

\begin{nth} One has
\beq\label{sil2}
\Si(\xi)=\xi(1+\tau)\,{\cal L}^2+{\cal M}^2, \qq\quad \tau=-\frac{\xi}{s^2}.
\eeq
\end{nth}

\vspace{5mm}
\nin{\bf Proof:} Using the relations in (\ref{defsigma2})  we get
\beq
\Si(\xi)=\sum_{l=0}^n \left(\sum_{k=0}^l S_{k,l-k}\right)\xi^l+\sum_{l=n+1}^{2n}\left(\sum_{k=l-n}^nS_{k,l-k}\right)\xi^l=\sum_{k,l=0}^n S_{k,l}\xi^{k+l}.\eeq
Expressing $S_{k,l}$ in terms of the $\la's$ we have to compute the following terms:
\beq
\sum_{k=0}^n\la_{2k-1}\xi^k\sum_{l=0}^n\la_{2l-1}\xi^l={\cal M}^2,\qq\qq \sum_{k=1}^n \la_{2(k-1)}\xi^k\sum_{l=0}^{n-1}\la_{2l}\xi^l=\xi{\cal L}^2,\eeq
and
\beq
-\frac 1{s^2}\sum_{k=1}^n\la_{2(k-1)}\xi^k\sum_{l=1}^n\la_{2(l-1)}\xi^l=\xi\tau{\cal L}^2.\eeq
Adding up ends up the proof. $\hfill\Box$

Let us compute the explicit form of the moments in terms of the parameters $m_k$ appearing in $A(\tht)$. To this end we will define, for the string $M=(m_1,m_2,\ldots,m_{2n-1})$ the symmetric functions 
$(M)_k$:
\beq
\prod_{k=1}^{2n-1}\,(1+\xi\,m_k)=\sum_{k=0}^{2n-1}\,\xi^k\,(M)_k,
\eeq
and we will prove

\begin{nth} The generating function of the moments is
\beq\label{Si2}
\Si(\xi)=(1-\xi)\,\prod_{k=1}^{2n-1}\,(1-\xi\,m_k),\qq\quad \xi\in {\mb C},
\eeq
giving the explicit formulae
\beq\label{tabSi2}\barr{cl}
k=0: & \qq \si_0=1\\[4mm]
1\leq k\leq 2n-1: &\qq \si_k=(-1)^k[(M)_k+(M)_{k-1}]\\[4mm] k=2n: & \dst\qq\si_{2n}=(M)_{2n}=\prod_{k=1}^{2n-1}\,m_k.\earr
\eeq 
\end{nth}

\vspace{5mm}\nin{\bf Proof:} Let us restrict ourselves to $\tau>0$. Relation (\ref{gfM2}) gives
\beq
{\cal M}=(1+\tau)^n\left\{\sum_{l=0}^{n-1}\,\left(\frac{\tau}{1+\tau}\right)^l\,(H)_{2l}+c\,\sum_{l=0}^{n-1}\,\left(\frac{\tau}{1+\tau}\right)^{l+1}\,(H)_{2l+1}\right\}.
\eeq
The coordinate change
\beq
\frac{\tau}{1+\tau}=\eta^2\qq\qq\qq 1+\tau=\frac 1{1+\eta^2},
\eeq
allows to write
\beq
{\cal M}=(1+\tau)^n\left\{\sum_{l=0}^{n-1}\,\eta^{2l}\,(H)_{2l}+\eta\,c\,\sum_{l=0}^{n-1}\,\eta^{2l+1}\,(H)_{2l+1}\right\}.
\eeq
Using the relations (\ref{evenH}) and (\ref{oddH}) given in Appendix A leads to
\beq
{\cal M}=(1+\tau)^n\Big\{\frac 12(1+\eta c){\cal H}(\tht,+\eta) +\frac 12(1-\eta c){\cal H}(\tht,-\eta) \Big\}
\eeq
Using relation (\ref{gfL2}), and after similar steps, one gets
\beq
-s(1+\tau){\cal L}=(1+\tau)^n\Big\{\frac 1{2\eta}(1+\eta c)\,{\cal H}(\tht,+\eta) -\frac 1{2\eta}(1-\eta c)\,{\cal H}(\tht,-\eta)\Big\}.
\eeq
The last step uses (\ref{sil2}): 
\beq
\Si(\xi)={\cal M}^2-\eta^2\Big(s(1+\tau){\cal L}\Big)^2=(1+\tau)^{2n}(1-\eta^2 c^2)\,{\cal H}(\tht,+\eta)\,{\cal H}(\tht,-\eta).
\eeq
The proof of (\ref{Si2}) follows from the relations
\beq
1-\eta^2\,c^2=\frac{1-\xi}{1+\tau}\qq\qq (1+\tau)^{2n-1}\,{\cal H}(\tht,+\eta)\,{\cal H}(\tht,-\eta)=\prod_{l=1}^{2n-1}(1-\xi\,m_l).
\eeq
Analytic continuation extends this result to $\xi \in{\mb C}$. Expanding $\Si(\xi)$  in powers of $\xi$ gives the relations in (\ref{tabSi2}). $\hfill\Box$ 

\vspace{5mm}\nin{\bf Remark:} It follows again that
\beq
\Si(1)=\sum_{l=0}^{2n}\,\si_l =0.\eeq

Let us conclude this section with

\begin{nth} One has the relation
\beq\label{comS}
\{S_+,S_-\}=2i\,\sum_{l=0}^{2n-1}\,(l+1)\si_{l+1}\,H^{2n-l-1}\,\Pf^{2l+1}.\eeq
\end{nth}

\nin{\bf Proof:} Extracting out from the bracket the 
$\phi$ dependence gives
\beq
\frac{\{S_+,S_-\}}{2i}=\frac 12\frac{\pt}{\pt \Pf}({\cal S}^2+{\cal T}^2)-\{{\cal S},{\cal T}\}.
\eeq
The first term in the right hand side gives
\beq
\sum_{l=0}^{2n-1}\,(l+1)\si_{l+1}\,H^{2n-l-1}\,\Pf^{2l+1}+\frac 1{s^2}\sum_{l=0}^{2n}(2n-l)\,\si_l\,H^{2n-l}\,\Pf^{2l+1},\eeq
so that the relation (\ref{comS}) will hold true if we can prove the relation
\beq\label{null2}
s^2\{{\cal S},{\cal T}\}=\sum_{l=0}^{2n}(2n-l)\,\si_l\,H^{2n-l}\,\Pf^{2l+1}.\eeq
Defining $\Psi^{2n}_{k+l}=H^{2n-k-l}\Pf^{2(k+l)-1}$ let us first compute the left hand member:
\beq\barr{l}\dst 
\{{\cal S},{\cal T}\}=\sum_{k,l=0}^n\frac{\Psi^{2n}_{k+l}}{H}\Big[(n-k)\la_{2k-1}\{H,\la_{2(l-1)}\Pi\}-(n-l)\la_{2(l-1)}\Pi\{H,\la_{2k-1}\}+\\[4mm]
\hspace{7cm}+H\{\la_{2k-1},\la_{2(l-1)}\Pi\}\Big].
\earr\eeq
Let us change, in the second sum, $l \to k$. Since we have
\beq
\{H,\la_{2l-1}\}=2\Pi\,\frac{\la'_{2l-1}}{A}=-\frac{2\Pi}{s^2}\,\la_{2(l-1)},\eeq
using relation (a) in (\ref{sd2}), one gets
\beq\barr{l}\dst 
s^2\{{\cal S},{\cal T}\}=\sum_{k,l}\frac{\Psi^{2n}_{k+l}}{H}\Big[2(n-k)\la_{2k-1}\frac{s^2\la'_{2(l-1)}}{A}\Pi^2+2(n-k)\la_{2(k-1)}\la_{2(l-1)}\Pi^2+\\[4mm]\dst 
\hspace{5.8cm}+2(n-k)\frac c{sA}\la_{2k-1}\la_{2(l-1)}\Pf^2+H\la_{2(k-1)}\la_{2(l-1)}\Big].\earr\eeq
Using $\dst \Pi^2=H-\frac{\Pf^2}{s^2}$ leads to
\beq\barr{l}\dst 
s^2\{{\cal S},{\cal T}\}=\sum_{k,l}\Big[(2n-2k+1)\la_{2(k-1)}\la_{2(l-1)}+2(n-k)\la_{2k-1}\frac{s^2\la'_{2(l-1)}}{A}\Big]\Psi^{2n}_{k+l}
\\[4mm]\dst +\sum_{k,l}2(n-k)
\Big[-\la_{2k-1}\frac{\la'_{2(l-1)}}{A}-\frac 1{s^2}
\la_{2(k-1)}\la_{2(l-1)}+\frac c{sA}\la_{2k-1}\la_{2(l-1)} \Big]\Psi^{2n}_{k+l+1}.\earr\eeq
Changing the summation index $l-1 \to l$, (recall that  $\la_{2n}=\la_{-2}=0$), we have
\beq
\sum_k\sum_{l=1}^n 2(n-k)\la_{2k-1}
\frac{s^2\la'_{2(l-1)}}{A}\Psi^{2n}_{k+l}=\sum_{k,l} 
2(n-k)\la_{2k-1}\frac{s^2\la'_{2l}}{A}.
\eeq
Collecting the terms which display a factor $A^{-1}$ we obtain
\beq
\sum_{k,l} 2(n-k)\frac{\la_{2k-1}}{A}\left(s^2\la'_{2l}-\la'_{2(l-1)}+\frac cs\la_{2(l-1)}\right)\Psi^{2n}_{k+l+1}=\sum_{k,l} 2(n-k)\la_{2k-1}\la_{2l-1}\Psi^{2n}_{k+l+1},\eeq
using the relations (b) in (\ref{sd2}). So we conclude to
\beq\barr{l}\dst 
s^2\{{\cal S},{\cal T}\}=\sum_{k,l}(2n-2k+1)\la_{2(k-1)}\la_{2(l-1)}\Psi^{2n}_{k+l}+\\[4mm]\dst
\hspace{4.5cm}+\sum_{k,l}2(n-k)\left(\la_{2k-1}\la_{2l-1}-\frac 1{s^2}\la_{2(k-1)}\la_{2(l-1)}\right)\Psi^{2n}_{k+l+1}.\earr\eeq
Let us now consider the right hand member of (\ref{null2}):
\beq
\sum_{L=0}^{2n}(2n-L)\si_L H^{2n-L-1}\Pf^{2L+1}.\eeq
Expressing the $\si_L$ as in (\ref{defsigma2}) and exchanging the summations we get
\beq
\sum_{k}\sum_{L=k}^{k+n}(2n-L)S_{k,L-k} H^{2n-L-1}\Pf^{2L+1}=\sum_{k,l}(2n-k-l)S_{k,l}\Psi^{2n}_{k+l+1}.\eeq
Let us recall that
\[
S_{k,l}=\la_{2(k-1)}\la_{2l}+\la_{2k-1}\la_{2l-1}-\frac 1{s^2}\la_{2(k-1)}\la_{2(l-1)}.\]
So we have a first piece
\beq
\sum_{k,l} 2(n-k)\left(\la_{2k-1}\la_{2l-1}-\frac 1{s^2}\la_{2(k-1)}\la_{2(l-1)}\right)\Psi^{2n}_{k+l+1},
\eeq
while in the second the change $l \to l-1$ leads to
\beq
\sum_{k,l}(2n-k-l)\la_{2(k-1)}\la_{2l}\Psi^{2n}_{k+l+1}=\sum_{k,l}(2n-2k+1)\la_{2(k-1)}\la_{2(l-1)}\Psi^{2n}_{k+l}.
\eeq
These two pieces prove (\ref{null2}), hence the Proposition. $\hfill\Box$

We can therefore conclude to:

\begin{nth} The set of observables
\beq
H,\qq\quad \Pf, \qq\quad S_+= e^{-i\phi}\,({\cal S}+ i{\cal T}),\qq\quad S_-=e^{i\phi}\,({\cal S}\,-i{\cal T}),\eeq
is indeed a Poisson algebra with
\beq
S_+S_-=\sum_{l=0}^{2n}\,\si_l\,H^{2n-l}\,\Pf^{2l}, \qq \{S_+,S_-\}=2i\,\sum_{l=0}^{2n-1}\,(l+1)\si_{l+1}\,H^{2n-l-1}\,\Pf^{2l+1}.
\eeq
\end{nth}

\subsection{Global aspects}
We have considered the metric and the SI hamiltonian 
\beq
g=A^2(\tht)\,d\tht^2+s^2\,d\phi^2, \qq H=\Pi^2+\frac{\Pf^2}{s^2},\qq\qq \Pi=\frac{\Pt}{A(\tht)},
\eeq 
where
\beq
A(\tht)=1+{\cal A}(\tht), \qq {\cal A}(\tht)=c\sum_{k=1}^{2n-1}\frac{e_k}{\sqrt{1-m_k\,s^2}}.
\eeq

Let us prove

\begin{nth} The metric constructed above is 
{\bf never} globally defined on $M={\mb S}^2$.
\end{nth}

\vspace{5mm}{\bf Proof:} The metric will be globally defined on ${\mb S}^2$ provided that ${\cal A}([0,\pi])\subset\,(-1,+1)$. This is not the case since
\beq
{\cal A}(\tht=0)=-{\cal A}(\tht=\pi)=S+e_{2n-1} \quad S=\sum_{k=1}^{2n-2}\,e_k.
\eeq
If $S$ is strictly positive, then ${\cal A}(\tht=0)\geq 1$. If $S=0$ and $e_{2n-1}$ is positive we have ${\cal A}(\tht=0)=1$. If $S=0$ and $e_{2n-1}$ is negative, then ${\cal A}(\tht=0)=-1$. If $S$ is strictly negative just reverse $\tht=0$ and $\tht=\pi$.

It cannot be Zoll since ${\cal A}(0)=-{\cal A}(\pi)$ cannot vanish. $\hfill\Box$

\vspace{5mm}
\nin{\bf\large This concludes the proof of Theorem 2}.
$\hfill\Box$

\vspace{3mm}
\nin{\bf Remarks:}
\brm
\item The difference between the case of extra integrals with even degrees and odd degrees is quite surprising. However one could already observe this phenomenon for the $(1,2)$ Koenigs case.

\item Since these metrics are not globally defined, they are of little interest. Nevertheless the formulae we gave for the geodesics in Section 3, for the case of extra integrals of odd degrees are still valid: one needs just to change everywhere the summations over 
$\{1,2,\ldots,2n\}$ into summations over $\{1,2,\ldots,2n-1\}$.
\erm

\section{An example: the  quartic case}
Since it was studied by Novichkov in \cite{No}, let us examine this case more closely.

\subsection{Our solution}
Our solution of the problem was obtained using the coordinates $(\tht,\,\phi)$. We have for hamiltonian
\beq\label{hamG}
H=\Pi^2+\frac{\Pf^2}{s^2}, \qq \Pi=\frac{P_{\tht}}{A}, \qq A=1+c\sum_{k=1}^3 \frac{e_k}{\sqrt{1-m_k\,s^2}}.
\eeq
The extra integrals are given by
\beq
S_1=\cos\phi\,{\cal S}+\sin\phi\,{\cal T},\qq\qq S_2=\cos\phi\,{\cal T}-\sin\phi\,{\cal S},\eeq
with
\beq
{\cal S}=H^2+\la_1\,H\,\Pf^2+\la_3\,\Pf^4,\qq {\cal T}=\Pi\,\Pf\Big(\la_0\,H+\la_2\,\Pf^2\Big),
\eeq
where
\beq
\left\{\barr{ll}\dst 
\la_0=-\frac 1s\Big((H)_1+c\Big), & \dst \quad\la_2=\frac 1{s^3}\Big((H)_3+c\,(H)_2+(H)_1+c\Big),\\[4mm]\dst
\la_1=-\frac 1{s^2}\Big((H)_2+c\,(H)_1+2\Big), & \dst\quad  
\la_3=\frac 1{s^4}\Big(c\,(H)_3+(H)_2+c\,(H)_1+1\Big).
\earr\right.\eeq
Later on we will need also the quadratic relations which follow from the conservation of $S_1^2+S_2^2$:
\beq\label{Scarre}\barr{crcl}
(a)\qq\qq & \la_0^2+2\la_1 & = & \si_1,\\[4mm]
(b)\qq\qq  & \dst \la_1^2-\frac{\la_0^2}{s^2}+2\la_0\la_2+2\la_3 & = & \si_2, \\[4mm]
(c)\qq\qq  & \dst \la_2^2-2\frac{\la_0\la_2}{s^2}+2\la_1\la_3 & = & \si_3,\\[4mm]
(d)\qq\qq  & \dst  \la_3^2-\frac{\la_2^2}{s^2} & = & \si_4.\earr
\eeq
Now let us state Novichkov's results.

\subsection{Novichkov results}
They are expressed in the coordinates $(x,\,\phi)$ with the hamiltonian
\beq\label{hamN}
H=\Pi^2+h_x^2\,\Pf^2, \qq\quad \Pi=h_x\,P_x, \qq\quad  h_x=D_x\,h(x).
\eeq
Up to slight changes, his extra integrals $(S_1,\,S_2)$ are constructed from
\beq
{\cal S}=H^2+l_1(x)\,H\,\Pf^2+l_3(x)\,\Pf^4, \qq {\cal T}=P_x\Pf(l_0(x)\,H+l_2(x)\,\Pf^2).
\eeq
Let us first state his main result and present a short check:

\begin{nTh}[Novichkov] The previously defined dynamical system is SI provided that $\ h\ $ be a solution of the first order non-linear ODE:
\beq\label{eqNo}
{\cal N}(h)\equiv 2P\,h\,h_x^3+\Big[-(h^2+2A_2)P'+A_5\Big]h_x^2-P^2=0,\eeq
where
\beq
P=A_3\,\cosh x+A_4\,\sinh x.
\eeq
\end{nTh}

\vspace{5mm}\nin{\bf Check:} The conservation of $S_1$ gives 
\beq
l'_1=-l_0, \qq l'_2=-l_3, \qq l'_3=-l_2,\eeq
as well as
\beq
l'_0-\frac{h_{xx}}{h_x}l_0=h_x^2,\qq l'_2-\frac{h_{xx}}{h_x}l_2=h_x^2\Big(h_x^2+l_1+\frac{h_{xx}}{h_x}l_0\Big).\eeq
An obvious consequence is $\ l''_2=l_2$ which gives
\beq
l_2(x)\equiv P(x)=A_3\,\cosh x+A_4\,\sinh x,\qq\qq l_3=-P'.\eeq
Integrating for $l_0$ one gets
\beq
l_0=(h+A_1)h_x \qq\Longrightarrow\qq l_1=-\frac 12(h+A_1)^2-A_2.\eeq
It remains to use the relation involving $l_2$, which produces a second order ODE 
\beq\label{No2} 
\Big((h+A_1)h_x^3+P\Big)h_{xx}+h_x^5+\Big(-\frac 12(h+A_1)^2-A_2\Big)h_x^3-P'\,h_x=0, \eeq
and this is nothing but the second order ODE (2.3.3) obtained in \cite{No}. Let us observe that we can take $A_1=0$ by a translation of $h$, since only $h_x$ appears in the hamiltonian. 

Then, by the construction of an integrating factor,  Novichkov reduces (\ref{No2}) to the first order ODE (\ref{eqNo}). $\hfill\Box$

\subsection{Connection relations}
They follow from a comparison of the hamiltonians (\ref{hamG}) and (\ref{hamN}) and read
\beq
h_x(\tht)=\frac 1s \qq\quad D_{\tht}\,x=\frac{A}{s}\qq\quad D_{\tht}\,h(\tht)=\frac A{s^2}.
\eeq
Integrating for $x(\tht)$ and $h(\tht)$ one obtains
\beq\label{xtheta} 
x(\tht)-x_0=\ln\left(\frac{s}{1+c}\right)+\sum_{k=1}^3\ln\frac{s}{(1+h_k(\tht)}, \qq h(\tht)=-\frac 1s((H)_1+c),
\eeq
where $h_k(\tht)=e_k\sqrt{1-m_k\,s^2}$. 

A comparison of the extra integrals gives two new relations
\beq\label{PPp}
P=\frac 1{s^4}\Big((H)_3+c(H)_2+(H)_1+c\Big) \qq P'=-\frac 1{s^4}\Big(c(H)_3+(H)_2+c(H)_1+1\Big).
\eeq

Now we are in position to connect both formulations. We have first
\beq
e^x=\frac{e^{x_0}}{M}\frac 1{s^4}\,E_-, \qq E_-=(1-c)\prod_{k=1}^3(1-h_k),\qq M=m_1m_2m_3,
\eeq
and
\beq
e^{-x}=e^{-x_0}\frac 1{s^4}\,E_+, \qq E_+=(1+c)\prod_{k=1}^3(1+h_k).
\eeq
Let us notice the useful relations
\beq\barr{lcl}\dst 
\frac 12(E_+ + E_-) & = & c(H)_3+(H)_2+c(H)_1+1,\\[4mm]\dst \frac 12 (E_+ - E_-) & = & (H)_3+c(H)_2+(H)_1+c.\earr\eeq
So, starting from
\beq
P=\alf\,e^x+\be\,e^{-x},\qq \alf=\frac{A_3+A_4}{2},\quad \be=\frac{A_3-A_4}{2},
\eeq
using the previous formulae for the exponentials, and with the help of the relations
\beq
\alf\frac{e^{x_0}}{M}+\be\,e^{-x_0}=0 \qq \& \qq \alf\frac{e^{x_0}}{M}-\be\,e^{-x_0}=-1,
\eeq 
we conclude that $P$ and $P'$ are indeed given by the relations already obtained in (\ref{PPp}).

So the various objects appearing in Novichkov's ODE (\ref{eqNo}) are, in our notations:
\beq
h=\la_0, \qq h_x=\frac 1s, \qq P=\frac{\la_2}{s},\qq P'=-\la_3.\eeq
As a side remark, let us point out that our solution gives a {\em parametric} solution of (\ref{eqNo}) in terms of the coordinate $\tht$.

Now we will check this ODE. We have 
\beq
\frac{{\cal N}(h)}{h_x^2}=2P\,h\,h_x-(h^2+2A_2)P'+A_5-\frac{P^2}{h_x^2},\eeq
which, translated in our notations, becomes:
\beq
\frac{{\cal N}(h)}{h_x^2}=2\frac{\la_0\la_2}{s^2}+(\la_0^2+2A_2)\la_3+A_5-\la_2^2.
\eeq
Using (\ref{Scarre})(c) leads to 
\beq
\frac{{\cal N}(h)}{h_x^2}=\la_3(\la_0^2+2\la_1)+2A_2\,\la_3+A_5-\si_3,
\eeq
and using (\ref{Scarre})(a), we end up with
\beq
\frac{{\cal N}(h)}{h_x^2}=(2A_2+\si_1)\la_3+A_5-\si_3,\eeq
which does vanish by the identification of parameters
\beq
A_2=-\frac{\si_1}{2} \qq\qq A_5=\si_3.\eeq

Let us explain now why there is no way to give an {\em explicit} solution to (\ref{eqNo}) keeping the coordinate 
$x$. The solution we obtained is {\em explicit} provided that one is using for coordinate $\tht$. Now, looking at the formula (\ref{xtheta}) for $x(\tht)$ it is clear that its  reciprocal function cannot be explicit.


\section{Conclusion}
Let us conclude with the following remarks:
\begin{itemize}
\item We have seen the importance of a ``good" choice of the coordinates in order to be able to solve explicitly the differential systems of SI systems. Unfortunately the choice of ``good" coordinates is not algorithmic.
\item We have proved the {\em existence} of a solution for the differential systems (\ref{sd1}) and (\ref{sd2}). However the problem of {\em uniqueness} is left open.
\item The main surprise of this article is probably that SI systems are not necessarily Zoll, even for metrics of revolution! This is particularly striking for the case of integrals of even degrees. Our conjecture that the converse is true, i. e. that any Zoll metric of revolution generates a SI system, remains an open problem.
\item In the approach of Matveev and Shevchishin 
\cite{ms}, one considers extra integrals having three different dependences with respect to the coordinate $\phi$:
\brm
\item A trigonometric dependence, considered in this work. For extra integrals of odd degree in the momenta we have obtained SI systems globally defined on ${\mb S}^2$.
\item A hyperbolic dependence. In the cubic case this choice led to no globally defined metric, so it does not seem very attractive to generalize it to higher degrees.
\item A quadratic dependence. This case was solved in \cite{Va3} for any degree of the extra integrals: it  leads to metrics globally defined either  on ${\mb R}^2$ or on ${\mb H}^2$ but never on ${\mb S}^2$.\erm
\item If one starts looking for a SI system with one Killing vector $\pt_{\phi}$ and a quadratic integral of the form
\beq
S=A(\tht,\phi)P_{\tht}^2+B(\tht,\phi)\,P_{\tht}\,\Pf+C(\tht,\phi)\,\Pf^2,
\eeq
one can prove that the only possible 
$\phi$-dependence of the various functions is, as considered in \cite{ms}, either trigonometric or hyperbolic or quadratic and {\em there is no other possibility}. This is Koenigs theorem \cite{Ko}. However, it is an open problem to ascertain whether this remains true for  SI systems with cubic and higher degree integrals. 
\item The study of the quantization of all of these models could be interesting albeit difficult. The conformally invariant quantization constructed in \cite{dlo} may play a prominent role.
\end{itemize}

\begin{appendices}

\section{Appendix A}
The functions $h_k(\tht)$ such that
\[\forall k\in\{1,2,\ldots,\nu\}: \qq\qq h_k(\tht)=e_k\,\sqrt{1-m_k\,s^2}\qq\quad e_k^2=1,\]
allow to define the functions $(H)_k(\tht)$ by the generating function

\beq\label{AppgenH}
{\cal H}(\tht,\xi)\equiv\prod_{k=1}^{\nu}(1+\xi\,h_k(\tht))=\sum_{k=0}^{\nu}\,(H)_k(\tht)\,\xi^k.\eeq

\begin{nth} The derivatives with respect to $\tht$ of the 
functions $(H)_k$ are given by
\beq\label{derH}
\forall k\in\{1,2,\ldots,\nu\}: \quad (H)'_{k}=k\,\frac cs\,(H)_k+(\nu-k+2)\,\frac cs\,(H)_{k-2}-\frac{A-1}{s}(H)_{k-1},
\eeq
provided that 
\beq
(H)_{-1}(\tht)\equiv 0, \qq\qq A(\tht)=1+c\,\sum_{k=1}^{\nu}\frac 1{h_k(\tht)}.\eeq
\end{nth}

\vspace{5mm}
\nin{\bf Proof:} Using the relations
\beq
h'_k=\frac cs\left(h_k-\frac 1{h_k}\right),
\eeq
we deduce
\beq
\frac{\pt_{\tht}{\cal H}}{\cal H} =\frac cs\left(\sum_{k=1}^{\nu}\frac{\xi h_k}{1+\xi h_k}-\sum_{k=1}^{\nu}\frac \xi{h_k(1+\xi h_k)}\right)=\frac cs\frac{\xi\pt_{\xi}{\cal H}}{\cal H}-\frac cs \sum_{k=1}^{\nu}\frac \xi{h_k(1+\xi h_k)}.
\eeq
The last sum is transformed according to
\beq\barr{lcl}\dst 
\sum_{k=1}^{\nu}\frac{\xi(1-\xi^2 h_k^2+\xi^2 h_k^2)}{h_k(1+\xi h_k)} & = & \dst \sum_{k=1}^{\nu}\frac{\xi(1-\xi h_k)}{h_k}+\xi^2\sum_{k=1}^{\nu}\frac{\xi h_k}{1+\xi h_k}\\[5mm] & = & \dst \xi\frac{A-1}{c}-\nu\,\xi^2+\xi^2\,\frac{\xi\pt_{\xi}{\cal H}}{\cal H}.\earr
\eeq
Hence we have obtained
\beq
\pt_{\tht}{\cal H}=\frac cs\Big(\xi\pt_{\xi}{\cal H}+\xi^2(\nu-\xi\pt_{\xi}){\cal H}\Big)-\xi\frac{A-1}{s}\,{\cal H}.
\eeq
Expanding in powers of $\xi$ gives (\ref{derH}).$\hfill\Box$

Let us mention also the useful relation
\beq\label{idH}
c\,(H)_{\nu-1}=(A-1)\,(H)_{\nu}.\eeq

Splitting in (\ref{AppgenH}) the even and the odd powers of $\xi$ gives 
\beq\label{evenH}
\frac 12({\cal H}(\tht,\xi)+{\cal H}(\tht,-\xi)) =\sum_{k=0}^{\nu}\,\xi^{2k}\,(H)_{2k}(\tht),\eeq
and 
\beq\label{oddH}
\frac 12({\cal H}(\tht,\xi)-{\cal H}(\tht,-\xi))=\sum_{k=0}^{\nu-1}\,\xi^{2k+1}\,(H)_{2k+1}(\tht).
\eeq

\end{appendices}


\begin{thebibliography}{2222}
\bibitem{Be} A. L. Besse, ``Manifolds all of whose geodesics are closed", Springer-Verlag, Berlin  
Heidelberg New-York (1978).
\bibitem{dlo} C. Duval, P. B. A. Lecomte and V, Ovsienko, {\sl Ann. Inst. Fourier (Grenoble)}  
{\bf 49} 1999-2029 (1978).
\bibitem{Ko} G. Koenigs, ``Sur les g\'eod\'esiques \`a int\'egrales quadratiques", a note appearing in 
``Le\c cons sur la th\'eorie g\'en\'erale des surfaces", G. Darboux, Vol 4,  368-404, Chelsea Publishing, (1972).
\bibitem{kk1}E. G. Kalnins, J. M. Kress and P. Winternitz, {\sl J. Math. Phys.},{\bf 43} 970-983 (2002).
\bibitem{kk2} E. G. Kalnins, J. M. Kress, W. Miller Jr and P. Winternitz, {\sl J. Math. Phys.} 
{\bf 44} (12) 5811-5848 (2003).
\bibitem{ms} V. S. Matveev and V. V. Shevchishin, {\sl J. Geom. Phys.}, {\bf 61} 1353-1377 (2011).
\bibitem{No} P. Novichkov, ``SI metrics on surfaces admitting integrals of degrees 1 and 4", Thesis (2015),
{\tt arXiv: 1805.10439 [math-ph]}.
\bibitem{Va1} G. Valent, {\sl Regul. Chaotic Dyn.}, 
{\bf 21} (5), 477-509 (2016). 
\bibitem{vds} G. Valent, C. Duval and S. Shevchishin, {\sl J. Geom. Phys.}, {\bf 87}, 461-481 (2015).
\bibitem{Va2} G. Valent, {\sl Lett. Math. Phys.}, {\bf 104}, 1121-1135 (2014).
\bibitem{Va3} G. Valent, {\sl Regul. Chaotic Dyn.}, {\bf 22} (4) 319-352 (2017).
\end{thebibliography}
\end{document}